\documentclass[iop]{emulateapj}

\usepackage{amssymb,amsmath,amsfonts,amsbsy}
\usepackage{color}
\usepackage{natbib}
\usepackage{enumerate}
\usepackage{graphicx}

\begin{document}

\title{Topology of Large-Scale Structures of Galaxies in Two Dimensions -- Systematic Effects}

\author{Stephen Appleby$^{a}$}\email{stephen@kias.re.kr}
\author{Changbom Park$^{a}$}
\author{Sungwook E. Hong$^{a}$}
\author{Juhan Kim$^{b}$}
\affiliation{$^{a}$School of Physics, Korea Institute for Advanced Study, 85
Hoegiro, Dongdaemun-gu, Seoul 02455, Korea}
\affiliation{$^{b}$Center for Advanced Computation, Korea Institute for Advanced Study, 85 Hoegiro, Dongdaemun-gu, Seoul 02455, Korea}

\begin{abstract}

We study the two-dimensional topology of the galactic distribution when projected onto two-dimensional spherical shells. Using the latest Horizon Run 4 simulation data, we construct the genus of the two-dimensional field and consider how this statistic is affected by late-time nonlinear effects -- principally gravitational collapse and redshift space distortion (RSD). We also consider systematic and numerical artifacts such as shot noise, galaxy bias, and finite pixel effects. We model the systematics using a Hermite polynomial expansion and perform a comprehensive analysis of known effects on the two-dimensional genus, with a view toward using the statistic for cosmological parameter estimation. We find that the finite pixel effect is dominated by an amplitude drop and can be made less than $1\%$ by adopting pixels smaller than $1/3$ of the angular smoothing length. Nonlinear gravitational evolution introduces time-dependent coefficients of the zeroth, first, and second Hermite polynomials, but the genus amplitude changes by less than $1\%$ between $z=1$ and $z=0$ for smoothing scales $R_{\rm G} > 9 {\rm Mpc/h}$. Non-zero terms are measured up to third order in the Hermite polynomial expansion when studying RSD. Differences in shapes of the genus curves in real and redshift space are small when we adopt thick redshift shells, but the amplitude change remains a significant $\sim {\cal O}(10\%)$ effect. The combined effects of galaxy biasing and shot noise produce systematic effects up to the second Hermite polynomial. It is shown that, when sampling, the use of galaxy mass cuts significantly reduces the effect of shot noise relative to random sampling.

\end{abstract}

\maketitle

\section{Introduction} 

The distribution of galaxies in the universe contains a wealth of information. Their clustering properties can be used for cosmological parameter estimation and to test the fundamental assumptions upon which the standard cosmological model is based. Conventional applications of galaxy data in cosmology have focused on the $N$-point correlation functions - effectively measuring the clustering as a function of scale. 

In parallel to the standard $N$-point analysis, an alternative approach has been developed that uses the topology of the galaxy distribution for cosmological parameter estimation \citep{Park:2009ja,Zunckel:2010eh,Speare:2013qma,Blake:2013noa} and also searching for non-Gaussian and modified gravity signals \citep{Matsubara:2000dg,Hikage:2006fe,Wang:2010ug,James:2011wm}. In this approach, the genus of the galactic density field is obtained from the data, and the conservation of genus with redshift provides a standard ruler or population by which one can reconstruct the `true' cosmology. The two methods extract similar information (both involve measuring the power spectrum); however, their sensitivity to systematic effects such as bias and redshift space distortions (RSDs) differ. In addition, the genus is sensitive to the phase structure of the density field on small scales, hence further information can be extracted from the genus curve by considering the evolution of its shape with redshift. This requires modeling of the genus shape in terms of cumulants of the density field in the nonlinear regime \citep{Pogosyan:2009rg,Gay:2011wz,Codis:2013exa}. 

The genus is perhaps the simplest and most commonly utilized measure of topology. It is a member of the Minkowski Functionals -- a set of scalar quantities that form a complete basis of valuations of a space \citep{Hadwiger}. Their origin can be traced to early works on integral geometry, and they have been utilized in cosmology for nearly three decades \citep{Gott:1986uz,Gott:1988rj,Ryden:1988rk,Gott:1988rj,1989ApJ...345..618M,1992ApJ...387....1P,1991ApJ...378..457P,Matsubara:1994we,1996ApJ...457...13M,Schmalzing:1995qn,2005ApJ...633....1P}
. In this paper, we focus exclusively on the two-dimensional genus of large-scale structure, which has been studied extensively in the literature \citep{1989ApJ...345..618M,1991MNRAS.250...75C,1992ApJ...387....1P,1993MNRAS.260..572C,2000ApJ...529..795C, 2001ApJ...553...33P, 2002ApJ...570...44H,1997ApJ...489..471C, Gott:2006za,2015ApJ...814....6W}.

The matter density field is three-dimensional in nature. However, in practice we do not always have access to full three-dimensional spatial information, for a number of reasons. One reason is that we measure the density field in redshift space, which is biased relative to its real space counterpart, due to both linear and nonlinear effects. In addition, cataloging large numbers of galaxies is typically achieved using photometric bands to estimate their redshift. Although recent advances in this field are impressive \citep{Bilicki:2013sza,Laigle:2016jxn}, photometric redshift estimation remains a dominant source of uncertainty. 

In this work, we study the two-dimensional genus of the galaxy distribution, tomographically binning galaxies in shells. By taking sufficiently thick redshift slices, we aim to mitigate both RSD and photometric redshift effects. The price we pay for these advantages is that the two-dimensional genus contains less information, relative to its three-dimensional counterpart. The Fourier modes parallel to the line of sight are effectively averaged in each redshift bin.

We use state-of-the-art simulations \citep{Kim:2015yma} to study the large-scale topological properties of the matter density field, using the distribution of dark matter particles and mock galaxy catalogs as tracers. The goal is a comprehensive study of the genus curve and how its properties are modified by various nonlinear and numerical processes. Many of these effects have been studied previously, both analytically and numerically: finite pixel effects have been expounded in \citet{1989ApJ...345..618M}, and gravitational evolution and RSD have been studied in \citet{Matsubara:1995dv} and \citet{1996ApJ...457...13M,2000astro.ph..6269M,2005ApJ...633....1P}. This work is a companion paper to \citet{Kim:2014axe}, where a similar analysis was performed for the three-dimensional genus.

The paper will proceed as follows. In section \ref{sec:covgrad}, we define the two-dimensional genus and calculate it for a simple Gaussian field. Following this, we describe how the genus is modified by various late-time effects, and use the latest Horizon Run 4 \citep{Kim:2015yma} N-body simulation to calculate departures from Gaussianity. We briefly describe the simulation data used in the analysis and proceed to study the consequences of pixel size, gravitational evolution, RSD, and shot noise in sections \ref{sec:pix}-\ref{sec:shot_noise}. We summarize in section \ref{sec:4}.

\section{two-dimensional Genus} 
\label{sec:covgrad}

Our aim is to study the two-dimensional genus of slices of the matter density field, using both dark matter particles and simulated galaxies as tracers. We begin with a discussion of the genus for a Gaussian random field $\delta(x,y,z)$ in three-dimensional space, with power spectrum $P_{\rm 3D}(k_{x},k_{y},k_{z})=P_{\rm 3D}(k_{\rm 3D})$ where $k_{\rm 3D} = \sqrt{k_{x}^{2}+k_{y}^{2}+k_{z}^{2}}$. We are interested in two-dimensional slices of this field, so we define a line of sight (without loss of generality $z$) and introduce a top-hat window function along this axis. The resultant two-dimensional field is then defined as the integral 

\begin{equation} \label{eq:1} \hat{\delta}_{\rm 2D}(x,y) =  \int \delta(x,y,z) F_{\rm z}(z) dz , \end{equation} 

\noindent with top hat 

\begin{equation} \label{eq:2} F_{\rm z}(z) = {1 \over \Delta_{z}}\left[\Theta \left( z-z_{\rm min} \right) - \Theta \left( z - z_{\rm max} \right) \right]  , \end{equation}

\noindent and bin thickness $\Delta_{z}=z_{\rm max}-z_{\rm min}$ along the line of sight. $\Theta(z)$ is the Heaviside step function. 

We also smooth the field in the $(x,y)$ plane perpendicular to the line of sight - for this purpose we adopt a Gaussian smoothing kernel. The final density field $\delta_{\rm 2D}$ is given by

\begin{equation}\label{eq:3} \delta_{\rm 2D}(x,y) = \int \delta(\tilde{x},\tilde{y},\tilde{z}) F_{\rm 2D}(|{\bf \tilde{r}} - {\bf r}|) F_{\rm z}(\tilde{z}) d^{2}\tilde{r} d\tilde{z} , \end{equation}

\noindent where ${\bf r} = (x,y)$ is the position vector in the two-dimensional space and 

\begin{equation} F_{\rm 2D}(|{\bf r}|) = {1 \over 2\pi R_{\rm G}^{2}} \exp \left[-{ r^{2} \over 2 R_{\rm G}^{2}} \right]  . \end{equation} 

\noindent where $R_{\rm G}$ is the two-dimensional Gaussian smoothing scale. The original three-dimensional field has power spectrum $P_{\rm 3D}(k_{\rm 3D})$, and the two-dimensional $P_{2D}(k_{\rm 2D})$ is related to $P_{\rm 3D}(k_{\rm 3D})$ as

\begin{eqnarray}\nonumber & &  P_{2D}(k_{\rm 2D}) = {L \over 2\pi} \int P_{\rm 3D}[(k_{\rm 2D}^{2} + k_{z}^{2})^{1/2}] \exp[-k_{\rm 2D}^{2}R_{\rm G}^{2}] , \\
\label{eq:4} & & \hspace{20mm} \times {\sin^{2} (k_{z}\Delta_{z}) \over k^{2}_{z}\Delta_{z}^{2}} dk_{z} , \end{eqnarray}

\noindent where $L^{3}$ is the total volume under consideration and $k_{\rm 2D} = \sqrt{k_{\rm x}^{2}+k_{\rm y}^{2}}$. For a Gaussian field, one can analytically calculate the expectation value of the genus curve, which is given by \citet{1989ApJ...345..618M}

\begin{equation} \label{eq:gauss}G_{2D}(\nu) = {1 \over (2\pi)^{3/2}} {\langle k_{\rm 2D}^{2} \rangle \over 2} \nu \exp[-\nu^{2}/2] , \end{equation}

\noindent where $\langle k_{\rm 2D}^{2} \rangle = \int k_{\perp}^{2} P_{2D}(k_{\perp}) d^{2} k_{\perp} /\int P_{2D}(k_{\perp}) d^{2}k_{\perp}$ is the square of the wavenumber $k_{\rm 2D}$ averaged over the two-dimensional power spectrum $P_{2D}(k_{\rm 2D})$. Thus, in the case of a Gaussian field, the amplitude of the genus is related to the shape of the power spectrum in the vicinity of the two-dimensional smoothing scale $R_{\rm G}$, and scales with the inverse square mean separation between peaks. The amplitude is a function of both $R_{\rm G}$ and $\Delta_{z}$.

The quantity $\nu$ is the density threshold used to define the excursion set for which we calculate the genus, normalized by the variance of the field. However, in what follows we choose not to calculate the genus directly in terms of the threshold overdensity $\nu$. Instead, we relate this quantity to the area fraction of the field that lies above a particular value. We define $\nu_{A}$ as

\begin{equation}\label{eq:afrac} f_{A} = {1 \over \sqrt{2\pi}} \int^{\infty}_{\nu_{A}} \exp[-t^{2}/2] dt , \end{equation}

\noindent where $f_{A}$ is the fractional area of the field above $\nu_{A}$. For example, when $\nu_{A} = 0$, we calculate the genus of the density field that occupies exactly half of the total area of the two-dimensional space, regardless of whether it is Gaussian-distributed. In the case of a Gaussian field, $\nu_{\rm A}$ is identical to $\nu$ within statistical fluctuations. More generally, there exists a bijective map between $\nu$ and $\nu_{\rm A}$: however, the $\nu_{\rm A}$ parameterization has been used to separately study non-Gaussianity beyond that in the one-point function \citep{1987ApJ...319....1G,1987ApJ...321....2W,1988ApJ...328...50M}. 

From an observational perspective, the galaxy samples that we typically use for cosmology are not homogeneous -- we introduce radial selection functions, and also line-of-sight effects such as RSDs must also be accounted for. Therefore, it will be convenient to project the density field onto shells centered on our location, and to calculate the genus on this surface rather than flat two-dimensional planes. Our analysis will closely follow that of \citet{Schmalzing:1997uc}, whose method we review for completeness.

For a two-dimensional field on an arbitrary surface $Q$, we can write the genus as \citet{Gott:1989yj}

\begin{equation} G_{2D} = {1 \over 2\pi} \int_{\partial Q} \kappa d \ell , \end{equation}

\noindent where $d\ell$ denotes the line element along the surface of $Q$ and $\kappa$ is the geodesic curvature. We calculate the surface density of the genus (that is, the genus per unit area) on the unit sphere, which is given by \citep{Schmalzing:1997uc}

\begin{equation} \label{eq:gen_d} g_{2D}(\nu) = {1 \over 4 \pi} \int_{Q} da \tilde{\delta}( \delta - \nu) { 2  \delta_{; 1}\delta_{; 2}\delta_{; 12}-\delta_{; 1}^{2}\delta_{; 22}- \delta_{; 2}^{2}\delta_{; 11} \over \delta_{; 1}^{2}+\delta_{; 2}^{2}} , \end{equation}

\noindent where $da$ now represents a surface element of the $2$-space $Q$, and covariant derivatives (first and second) are denoted with a semicolon. The indices $1,2$ represent an arbitrary coordinate system covering $Q$ and $\tilde{\delta}(x)$ denotes the Dirac delta function. 

To calculate the genus, our first step will be to pixelate the sphere. When doing so, we must calculate the first and second derivatives of $\delta$ at each pixel center. We also approximate the delta function as  

\begin{equation}\label{eq:delta} \tilde{\delta}(\delta-\nu) \simeq {1 \over \epsilon} {\bf I}_{(-\epsilon/2,+\epsilon/2)}(\delta-\nu) , \end{equation}

\noindent where ${\bf I}_{A}(x) = 1$ for $x \in A$ and ${\bf I}_{A}(x) = 0$ otherwise. The value of $\epsilon$ is chosen to be small, and the final result should be independent of its value. However, taking $\epsilon$ too small creates a sparse set of pixels that satisfy $\delta-\nu \in (-\epsilon/2,+\epsilon/2)$, generating numerical noise. If $\epsilon$ is too large, then the delta function is not well-approximated by ($\ref{eq:delta}$). Based on numerical testing of the stability of the genus, we take $\epsilon = 2.0 \times 10^{-3}(\delta_{\rm max} - \delta_{\rm min})$ in what follows, where $\delta_{\rm min, max}$ are the minimum/maximum values of the density field on the sphere. 

The discretized equivalent of equation ($\ref{eq:gen_d}$) is 

\begin{equation}\label{eq:gen_pix_2} g_{\rm 2D} \simeq {1 \over N_{\rm pix}}\sum_{i=1}^{N_{\rm pix}} \tilde{\delta}^{(i)} { 2  \delta^{(i)}_{; 1}\delta^{(i)}_{; 2}\delta^{(i)}_{; 12}-(\delta^{(i)}_{; 1})^{2}\delta^{(i)}_{; 22}- (\delta^{(i)}_{; 2})^{2}\delta^{(i)}_{; 11} \over (\delta^{(i)}_{; 1})^{2}+(\delta^{(i)}_{; 2})^{2}} , \end{equation}

\noindent where $\delta^{(i)}_{; 1}$ (for example) represents a discrete approximation to the full covariant derivative and $N_{\rm pix}$ is the total number of pixels. On non-flat $Q$-surfaces, it is practical to calculate covariant derivatives in Fourier space. We use the standard angular coordinate system 

\begin{equation} ds_{2}^{2} = d\theta^{2} + \sin^{2}(\theta) d\phi^{2} . \end{equation}

\noindent Coordinate labels $(1,2)$ in equations ($\ref{eq:gen_d}$) and ($\ref{eq:gen_pix_2}$) correspond to $\theta,\phi$.

For a Gaussian field, we can relate the amplitude of the genus to the angular power spectrum ${\cal C}_{\ell}$ as \citet{Schmalzing:1997uc}

\begin{eqnarray} \label{eq:gg} & &  g_{\rm 2D} = {1 \over 2(2\pi)^{3/2}} {\sigma_{1}^{2} \over \sigma_{0}^{2}} \nu \exp[-\nu^{2}/2] , \\
\label{eq:sig0} & & \sigma_{0}^{2} = \sum_{\ell=1}^{\infty} (2\ell+1) {\cal C}_{\ell} , \\
\label{eq:sig1} & & \sigma_{1}^{2} = \sum_{\ell=1}^{\infty} (2\ell+1) \ell (\ell + 1)  {\cal C}_{\ell} . \end{eqnarray} 

\begin{figure}
  \includegraphics[width=0.45\textwidth]{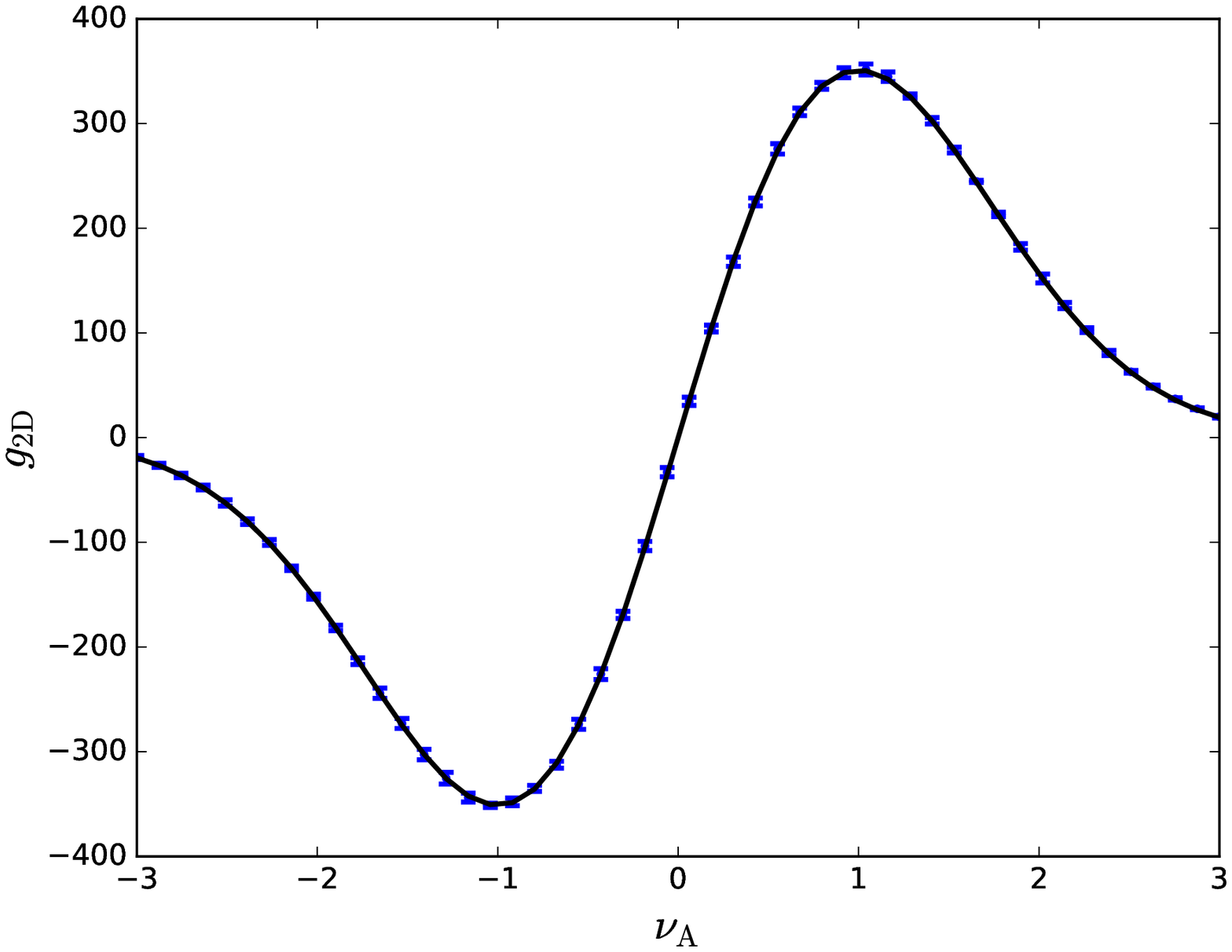}\\
  \includegraphics[width=0.45\textwidth]{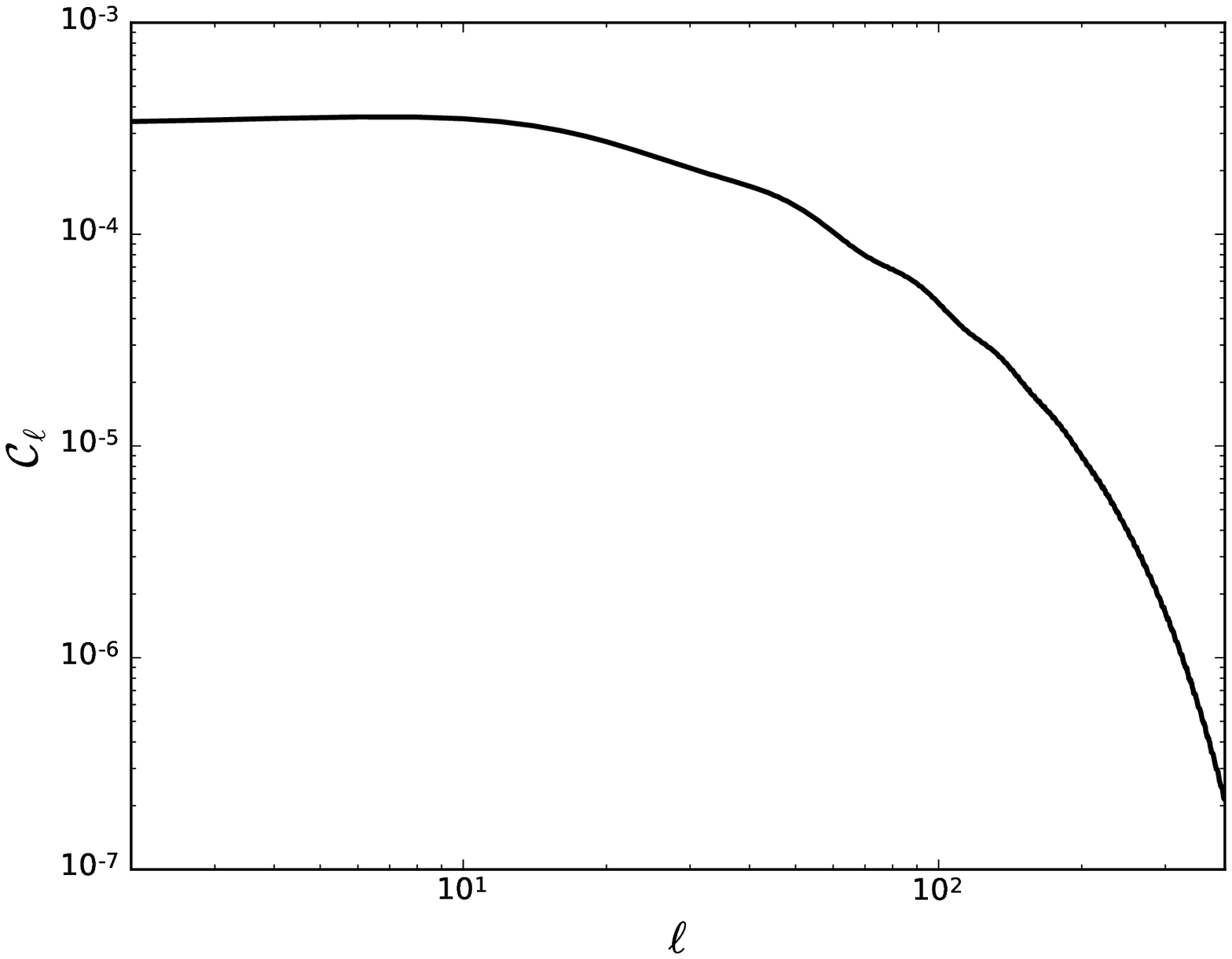}\\
  \caption{A genus curve for a Gaussian random field drawn from a sample angular power spectrum $C_{\ell}$, exhibited in the lower panel. The blue points and error bars were obtained by generating $N=200$ realizations of a Gaussian random field on the sphere and using our pixelated method to reconstruct the genus. The solid black line is the analytic prediction, equation ($\ref{eq:gg}$).}
  \label{fig:1}
\end{figure}

In Figure\,\ref{fig:1} (top panel), we exhibit the genus curve for a Gaussian realization of a field drawn from a $\Lambda$CDM power spectrum. The blue points are the values calculated using equation ($\ref{eq:gen_pix_2}$), and the black solid line is the theoretical expectation value equation ($\ref{eq:gg}$). The bottom panel exhibits the angular power spectrum of the Gaussian field on the unit sphere. The shape of the genus curve is completely fixed, and only the amplitude carries information. 

From inspection of equations ($\ref{eq:gg}-\ref{eq:sig1}$), it is clear that the amplitude of the genus is a measure of ratios of cumulants of the density field. As we are measuring ratios of the power spectrum, the amplitude will be insensitive to both linear bias and the linear growth rate. Despite this, one can still extract cosmological information from the genus amplitude by comparing its value in different redshift shells. Specifically, one can use the fact that the genus of the dark matter density field should be a conserved quantity when smoothed on large scales. However, if we choose an incorrect cosmology, then the physical smoothing scale that we adopt will systematically evolve with redshift, along with the measured surface area of the shell. The genus is sensitive to both of these quantities, and the two effects (incorrect smoothing scale and surface area) will not exactly cancel because the field is not scale-invariant. Hence, by measuring the genus in different redshift shells, we can reconstruct cosmological parameters by minimizing the evolution of the genus.

\section{Genus of a Non-Linear density field -- Horizon Run 4} 

Having completed our discussion on Gaussian random fields, we now consider a discrete point distribution of dark matter obtained from the Horizon Run 4 (HR4) simulation. The density field, obtained by smoothing the point distribution, is now non-Gaussian due to a number of nonlinear effects - chiefly nonlinear gravitational clustering. Before continuing, we briefly describe the simulation.

Horizon Run 4 \citep{Kim:2015yma} is the latest data release from the Horizon Run project\footnote{http://sdss.kias.re.kr/astro/Horizon-Runs}. It is a dense, cosmological-scale $N$-body simulation that gravitationally evolved $N=6300^{3}$ particles in a $V=(3150 \, {\rm Mpc/h})^{3}$ volume box. The simulation uses a modified version of GOTPM code and initial conditions are calculated using second-order Lagrangian perturbation theory, taking into account the experiments on the impact of initial conditions by \citet{L'Huillier:2014dpa}. The cosmological parameters used are given in Table \ref{tab:1}, and details of the simulation can be found in \citet{Kim:2015yma}. We will predominantly make use of snapshot data at $z=0,0.3,0.7,1.0,4.0$. In each snapshot box, we generate spheres of radius $d_{\rm cm}\sim 738 \, {\rm Mpc/h}$. With this choice, we can embed $n=8$ non-overlapping shells into the simulation box. 

We study two distinct density fields constructed from different point catalogs - the dark matter particle data and a mock galaxy catalog constructed in \citet{Hong:2016hsd}. We thin the particle data by a factor of four in each dimension - for our choice of pixel size the resulting number density is sufficient to minimize shot noise. The thinning procedure involves selecting every fourth particle along each dimension in the Lagrangian coordinates $(i,j,k)$, which outputs $(6300/4)^{3}$ particles. Here, $i,j,k$ are integer values. The initial particles are positioned at each grid point in the Lagrangian coordinates and they are perturbed to construct the desired density fluctuations, which are consistent with the input linear matter power spectrum. Therefore, to select fractions of particles from the total number, we have used their Lagrangian positions. Mock galaxies are assigned by the most bound halo particle-galaxy correspondence scheme. Survival time of satellite galaxies after merger is calculated by adopting the merger timescale model described in \citet{Jiang:2007xd}.

As we are now considering a non-Gaussian field, one can expect the genus curve to change in both amplitude and shape. In the mildly nonlinear regime, nonlinear effects have been analytically calculated and their effect on the genus shape is well-described by a low-order Hermite polynomial expansion \citep{2000astro.ph..6269M}. In the following sections, we typically calculate the residual of two genus curves based upon different assumptions - we define this quantity as $\Delta g_{\rm 2D}$. The quantity $\Delta g_{\rm 2D}$ will be defined differently in each of the following sections, according to the specific systematic quantity that we wish to study. For example, in section \ref{sec:RgDel} we analyze departures of the genus curve from its Gaussian limit; in this case, we define $\Delta g_{\rm 2D}$ as the difference between the measured genus curve and its Gaussian expectation ($\ref{eq:gg}$). In section \ref{sec:GC} we study the evolution of the genus with redshift; in that case $\Delta g_{\rm 2D}$ will denote the difference between the genus curve at high and low-z. We clearly define $\Delta g_{\rm 2D}$ at the beginning of each section.

Theoretically, we model $\Delta \hat{g}_{\rm 2D}$ using a Hermite polynomial expansion

\begin{equation}\label{eq:fit} \Delta \hat{g}_{\rm 2D}(\nu_{\rm A},a_{0-4}) =  A\exp[-\nu_{\rm A}^{2}/2] \sum_{i=0}^{4} a_{i} H_{\rm i}(\nu_{\rm A}) , \end{equation} 

\noindent with constant $a_{i}$, where $A$ is the amplitude calculated from equation ($\ref{eq:gg}$) 

\begin{equation} A = {1 \over 2(2\pi)^{3/2}} {\sigma_{1}^{2} \over \sigma_{0}^{2}} , \end{equation}

\noindent and the Hermite polynomials are given by 

\begin{eqnarray} & & H_{0}(x) = 1 , \\ 
& & H_{1}(x)= x , \\ 
& & H_{2}(x)= x^{2}-1 , \\ 
& & H_{3}(x)= x^{3}-3x , \\ 
& & H_{4}(x)= x^{4} - 6 x^{2}+3 .
\end{eqnarray}

\noindent We calculate $\Delta g_{\rm 2D}$ from the data for a set of $n=8$ independent shells within the Horizon Run box and then perform a $\chi^{2}$ minimization of  

\begin{equation}\label{eq:chi2} \chi^{2} = \sum_{\rm i=1}^{n_{\rm g}} {[ \Delta g_{\rm g, 2D}(\nu_{\rm A, i}) - \Delta\hat{g}_{\rm 2D}(\nu_{\rm A, i},a_{0-4})]^{2} \over \epsilon^{2}_{\rm i}} , \end{equation} 

\noindent to find the best fit $a_{0-4}$ values. Here, $i$ subscripts denote the $n_{\rm g}=50$ linearly spaced values of $\nu_{\rm A}$ at which we calculate the genus $g_{\rm 2D, i}$, and $\epsilon^{2}_{\rm i}$ is the population mean uncertainty of $\Delta g_{\rm 2D}(\nu_{\rm A, i})$, calculated using the eight distinct shells in the HR4 box. Specifically, $\epsilon_{\rm i}^{2}$ is obtained using a simple student t-test as $\epsilon_{\rm i} = t_{\alpha/2} s_{\rm i}/\sqrt{n}$, where $s_{\rm i}$ is the standard deviation of $\Delta g_{\rm 2D}(\nu_{\rm A, i})$, $n=8$, and the $\alpha$ subscript denotes the confidence limit chosen. We adopt the $95\%$ percentile for $t_{\alpha/2}$. We use the package COSMOMC \citep{Lewis:2013hha,Lewis:2002ah} as a generic sampler for the minimization. In applying this minimization procedure, we have assumed that the noise is Gaussian distributed, which is an increasingly poor approximation in the rare event tails of the distribution. When applying the genus statistic to data, accurate theoretical modeling of the statistical uncertainty must be undertaken. This will be the subject of future work.

\begin{table}
\begin{center}
%\captionof{table}{}\label{tab:1} 
 \begin{tabular}{||c | c ||}
 \hline
 Parameter & Fiducial Value \\ [0.5ex] 
 \hline\hline
 $\Omega_{\rm m0}$ & $0.26$   \\ 
 \hline
 $\Omega_{\Lambda}$ & $0.74$   \\ 
 \hline
 $n_{\rm s}$ & $0.96$   \\ 
 \hline
 $\sigma_{8}$ & $0.794$   \\
 \hline
 $\Delta$ & $0.02$   \\
 \hline
 $\theta_{\rm G}$ & $27.5$   \\
 \hline
 $z_{\rm min}$ & $0.25$   \\
 \hline
 $N_{\rm s}$ & $1024$   \\
 \hline
 $N_{\rm pix}$ & $12 \times N_{\rm s}^{2}$   \\
 \hline
\end{tabular}\label{tab:1}
%\caption*{Fiducial parameters used in the Horizon Run 4 simulation, and the parameters used to calculate the genus in this work. $\Delta$ is the redshift thickness of the spherical shells, $\theta_{\rm G}$ the angular smoothing scale (in radians) applied to the unit sphere. $z_{\rm min}$ is the redshift that defines the inner boundary of the shell, $N_{\rm pix}$ is the number of pixels used to discretize the sphere.}
\caption{Fiducial parameters used in the Horizon Run 4 simulation, and the parameters used to calculate the genus in this work. $\Delta$ is the redshift thickness of the spherical shells, $\theta_{\rm G}$ the angular smoothing scale (in arcminutes) applied to the unit sphere. $z_{\rm min}$ is the redshift that defines the inner boundary of the shell, $N_{\rm pix}$ is the number of pixels used to discretize the sphere.}
\end{center} 
\end{table}

\subsection{Results}
\label{sec:3}

We select eight locations in the $z=0$ snapshot data as observer points, separated such that we measure non-overlapping regions, and then construct shells between two redshift limits $z_{\rm min}=0.25$ and $z_{\rm max}=0.27$ (that is, shells of width $\Delta = 0.02$). Here, $z_{\rm min,max}$ are redshifts relative to the observer, and $z_{\rm mid}=(z_{\rm min}+z_{\rm max})/2$ denotes the approximate redshift of the center of the shell. Because we use snapshot data, $(z_{\rm min},z_{\rm max})$ are not true redshifts. However, the $z=0$ snapshot that we predominantly use is a reasonable approximation to lightcone data for small $z$.

We fix the total number of pixels as $N_{\rm pix}=12 \times N_{\rm s}^{2}$, where $N_{\rm s}=1024$, and pixelate the sphere according to the HEALPix\footnote{http://healpix.sourceforge.net} equal area scheme. The sphere is initially decomposed into twelve curvilinear quadrilaterals, and then subsequent subdivisions are made by dividing each pixel into four new ones until the desired pixel resolution is achieved. Details of the procedure can be found in \citet{Gorski:2004by}.

We use the HR4 cosmological parameters to relate the comoving distance along the line of sight $d_{\rm cm}(z)$ and the redshift $z$, and then bin all simulated galaxies that lie within the redshift range $(z_{\rm min},z_{\rm max})$ according to their nearest pixel center. Note that, when we perform the same calculation using galaxy data, no cosmological information is required at this stage. Cosmological dependence will only arise when we wish to compare the genus in different redshift shells, or when we wish to relate redshift and real space quantities. 

After binning the galaxies or dark matter particles, we obtain a series of pixels containing $n_{\rm i}$ points, where $i$ is the pixel identifier. We calculate the average $\bar{n}$ over the entire pixel range and define $\delta_{\rm i} = (n_{\rm i} - \bar{n})/\bar{n}$. We then perform two-dimensional Gaussian smoothing of $\delta_{\rm i}$ on the unit sphere using angular scale $\theta_{\rm G}$, which is related to the physical scale at the shell location as $R_{\rm G}\simeq d_{\rm cm}(z_{\rm mid}) \theta_{\rm G}$. We typically take $\theta_{\rm G} = 27.5 \,{\rm arcmin}$ in what follows, corresponding to a physical scale of $R_{\rm G} = 5.9 \, {\rm Mpc/h}$. 

To calculate the covariant derivatives of the field, we use HEALPix to first Fourier transform the field, and then calculate the derivatives using recursion relations between the Legendre polynomials. 

\begin{figure}
  \includegraphics[width=0.45\textwidth]{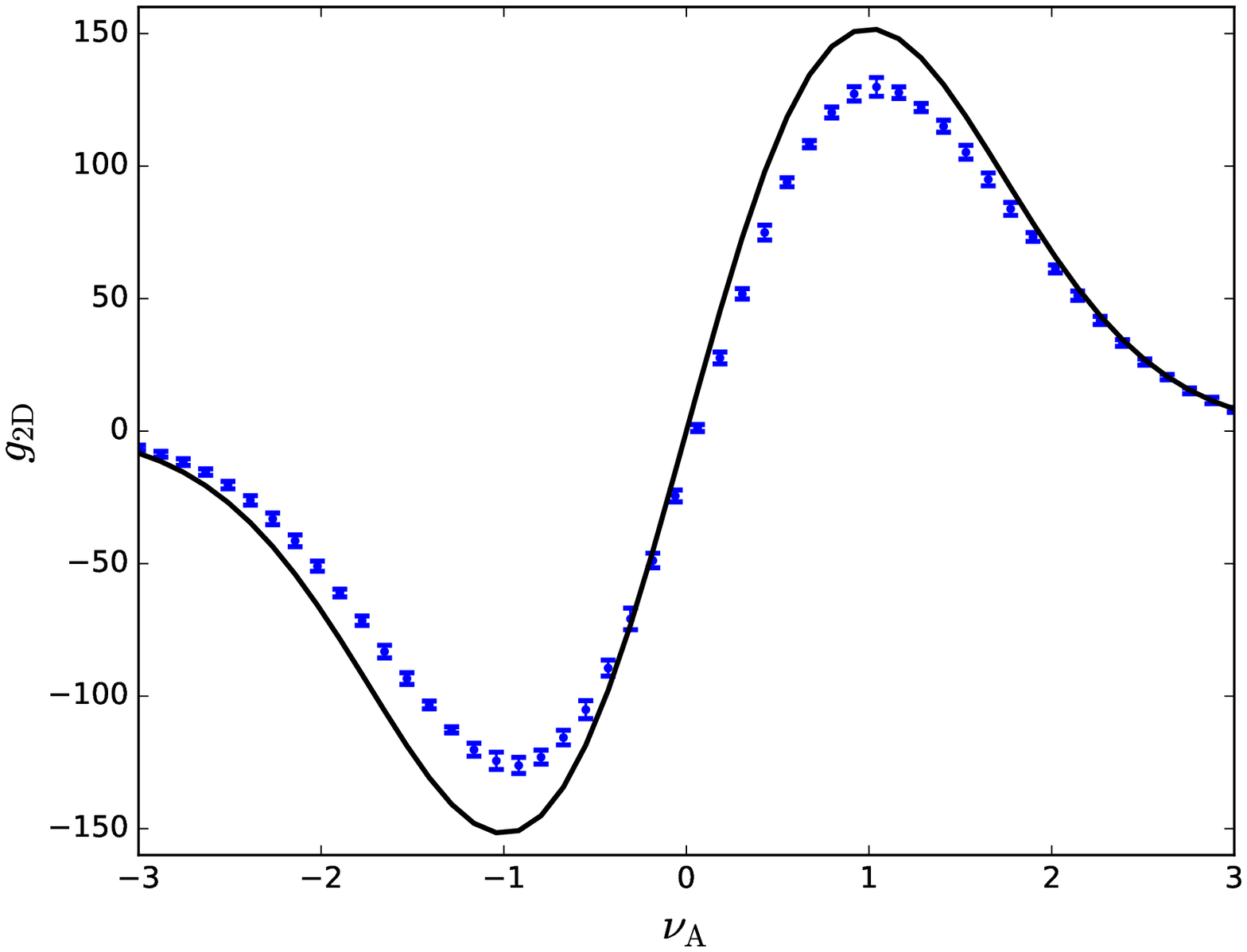}\\
  \includegraphics[width=0.45\textwidth]{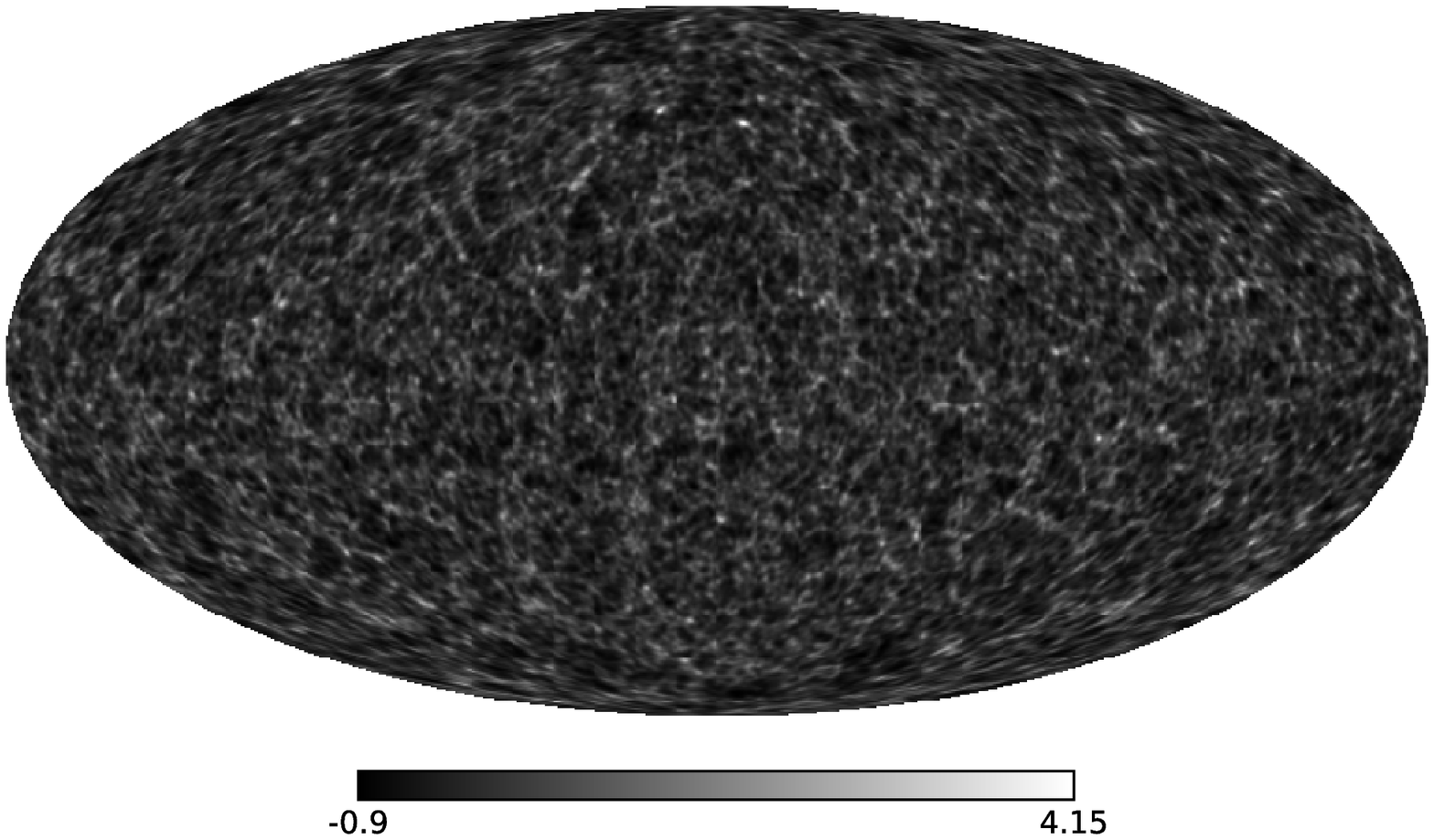}
  \caption{(Top panel) The dark matter genus curve as a function of $\nu_{\rm A}$. The points and error bars represent the sample mean and standard deviation of eight distinct shells generated on the $z=0$ HR4 snapshot data. The black solid line is the Gaussian expectation value based on the angular power spectrum of the two-dimensional density field. (Bottom panel) A Mollweide projected density map of a shell of the Horizon Run 4 dark matter density field, smoothed with our fiducial parameters $\Delta = 0.02$, $\theta_{\rm G} = 27.5\, {\rm arcmin}$.}
  \label{fig:2}
\end{figure}

Our baseline parameters are $\theta_{\rm G} = 27.5 \, {\rm arcmin}$ and $\Delta=0.02$; however, we will vary these in the following section. We exhibit the dark matter genus curve as a function of $\nu_{\rm A}$ in Figure\,\ref{fig:2}. The blue points and error bars represent the mean and standard deviation of the eight samples, and the black solid line denotes the genus curve constructed from the Gaussian expectation value ($\ref{eq:gg}$) using the angular power spectrum calculated from the data. 

The discrepancy between the Gaussian curve and the data indicates that our fiducial parameters are probing the nonlinear regime. The most striking departure from Gaussianity is an amplitude decrease, known as gravitational smoothing \citep{1989ApJ...345..618M}. However, the shape is also modified in an asymmetric manner. We discuss these effects further in the following section.

\subsection{Dependence on smoothing scales $\Delta$ and $\theta_{\rm G}$} 
\label{sec:RgDel}

Before studying gravitational and systematic effects, we begin by considering the dependence of the genus on the two smoothing scales $\Delta$ and $\theta_{\rm G}$. The dependence on $\theta_{\rm G}$ is crucial for cosmological parameter estimation, and the shell thickness $\Delta$ can also potentially contain cosmological information.  

In the Gaussian limit, the genus amplitude will depend quadratically on $\theta_{\rm G}$. Although we are considering the nonlinear regime, we still expect growth of the amplitude as we decrease the smoothing scale $\theta_{\rm G}$, by virtue of the increasing number of structures being resolved. 

Increasing $\Delta$ will `Gaussianize' the field along the line of sight, in the sense that when $\Delta$ is larger than the typical scale of bound structures we begin to stack structures along the line of sight. Stacking Fourier modes with correlated phases (as is the case in the late Universe) will have the effect of randomizing the phases. Mathematically, one can simply state that, as we increase $\Delta$, we are increasing the size of the real space top hat function. In Fourier space, we approach a delta function kernel when integrating over the modes parallel to the line of sight. All information in these modes is lost in the thick shell limit.

To examine sensitivity of $g_{\rm 2D}$ to $(\theta_{\rm G},\Delta)$, we define the function 

\begin{equation}\label{eq:dg_fid} \Delta g_{\rm 2D} = g_{\rm 2D, HR4}(\nu_{\rm A}) - g_{\rm 2D, G}(\nu_{\rm A}) , \end{equation}

\noindent where $g_{\rm 2D, HR4}(\nu_{\rm A})$ is the genus extracted from the HR4 data and $g_{\rm 2D, G}(\nu_{\rm A})$ is the Gaussian expectation value ($\ref{eq:gg}$), and the angular power spectrum is constructed from the data. We use both dark matter particle and mock galaxy data in this section, and apply the expansion ($\ref{eq:fit}$) to ($\ref{eq:dg_fid}$). The resulting coefficients $a_{0-4}$ represent amplitude and shape departures from Gaussianity.  

\begin{figure}
  \includegraphics[width=0.45\textwidth]{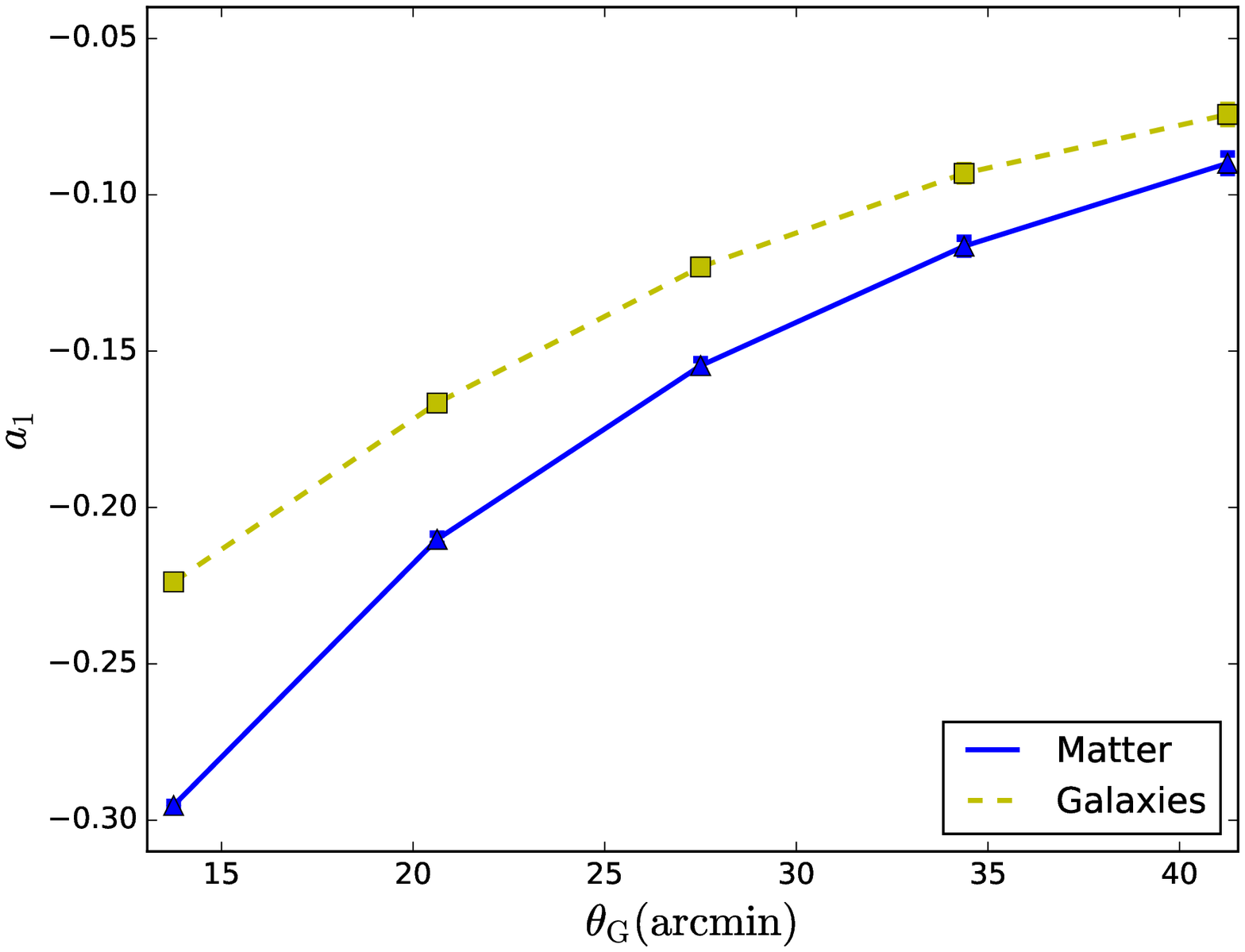}\\
  \includegraphics[width=0.45\textwidth]{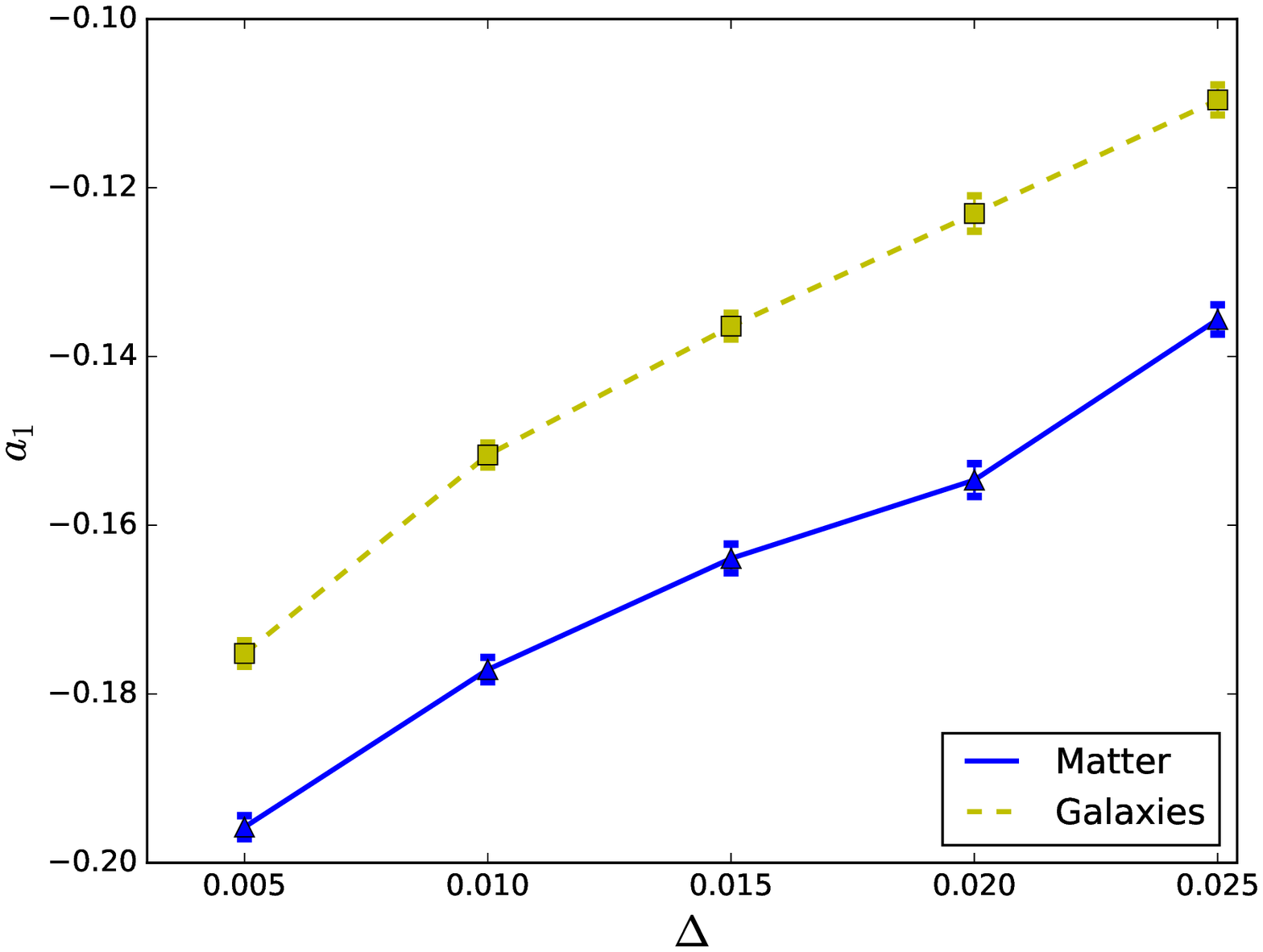}
  \caption{(Top Panel) The departure of the genus amplitude from its Gaussian limit ($a_{1}$) as a function of $\theta_{\rm G}$ in arcminutes, for our eight fiducial shells in the $z=0$ snapshot box. The blue triangles denote dark matter and yellow squares the simulated galaxy catalog. (Bottom Panel) The same quantity as a function of the shell thickness $\Delta$. The amplitude of the simulated density field is lower than the Gaussian limit, a well-known effect. One finds a slow approach to Gaussianity with increasing $\Delta$ and $\theta_{\rm G}$.}
  \label{fig:3}
\end{figure}

We first fix $\Delta=0.02$ and vary $\theta_{\rm G}$, and then fix $\theta_{\rm G} = 27.5 \, {\rm arcmin}$ and vary $\Delta$. The amplitude coefficient $a_{1}$ is exhibited as a function of $\theta_{\rm G}$ and $\Delta$ in Figure\,\ref{fig:3}. The blue triangles/solid line represent the amplitude obtained from the dark matter data, and the yellow squares/dashed line represents the mock galaxy catalog. The $a_{\rm 1}$ parameter is negative for the entire $(\theta_{\rm G},\Delta)$ range considered - this is due to gravitational smoothing \citep{1989ApJ...345..618M,1991ApJ...378..457P,2005ApJ...633....1P}.  The genus amplitude of the matter density field exhibits a stronger shift from the Gaussian limit than the galaxy distribution. This amplitude drop is not related to the linear bias between the samples - in fact, the genus amplitude is insensitive to (constant) bias factors. We note the approach to Gaussianity $a_{1} \to 0$ with increasing smoothing scale $\theta_{\rm G}$. 

When we fix $\theta_{\rm G} = 27.5 \, {\rm arcmin}$ and vary $\Delta$ over the range $0.0025 < \Delta < 0.025$, we also find a slow approach to Gaussianity $a_{1} \to 0$ with increasing $\Delta$. This is consistent with our expectations when stacking modes along the line of sight. When considering photometric galaxy data, we must fix the width of the tomographic bins such that they are wider than the typical photometric redshift uncertainty of the catalog. Current state-of-the-art in the field will force us to choose $\Delta > 0.02$, which corresponds to a physical scale of $\sim 60 {\rm Mpc/h}$. Even on these large scales, the genus can exhibit departures from Gaussianity at low redshift.  

\begin{figure}
  \includegraphics[width=0.45\textwidth]{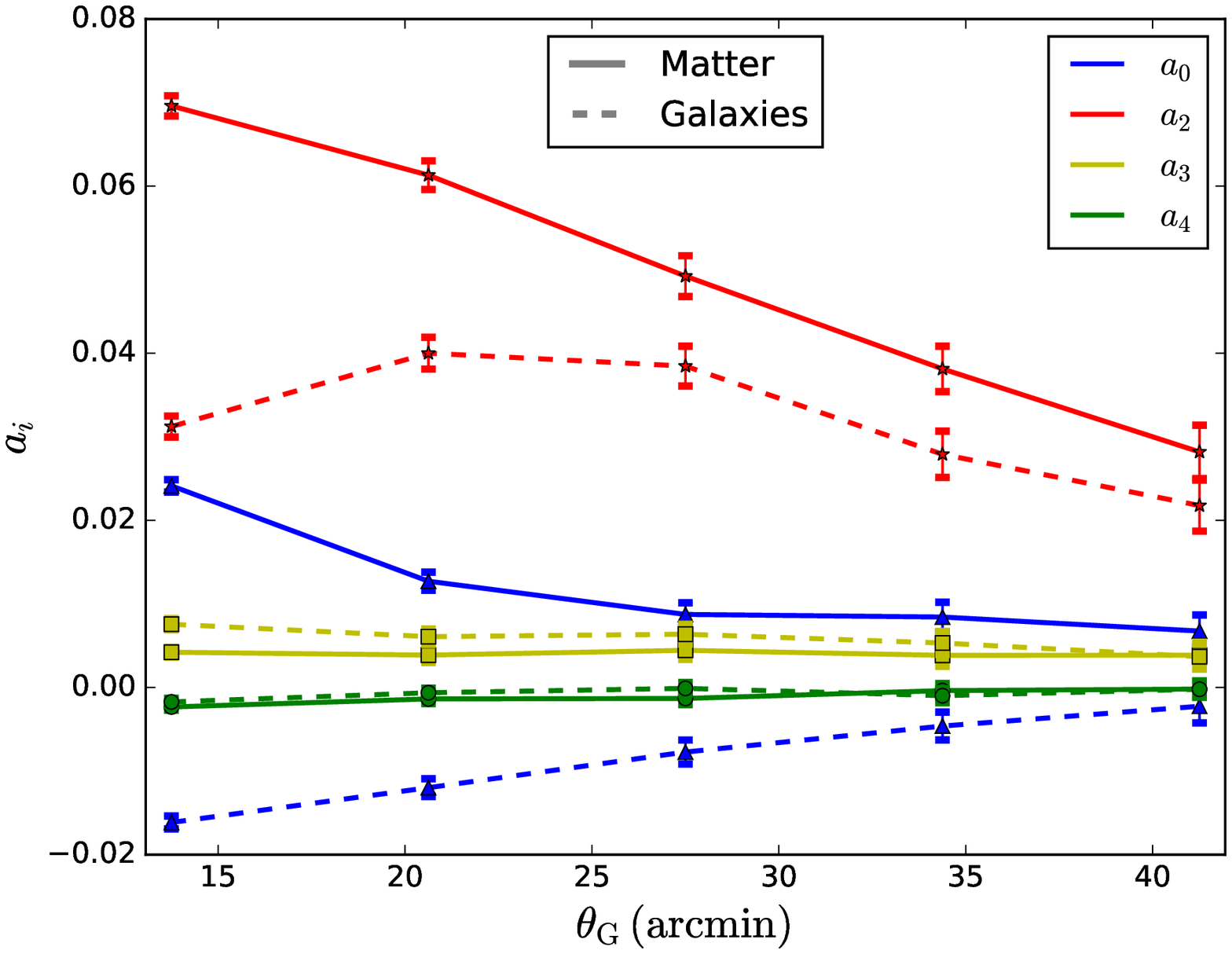}\\
  \includegraphics[width=0.45\textwidth]{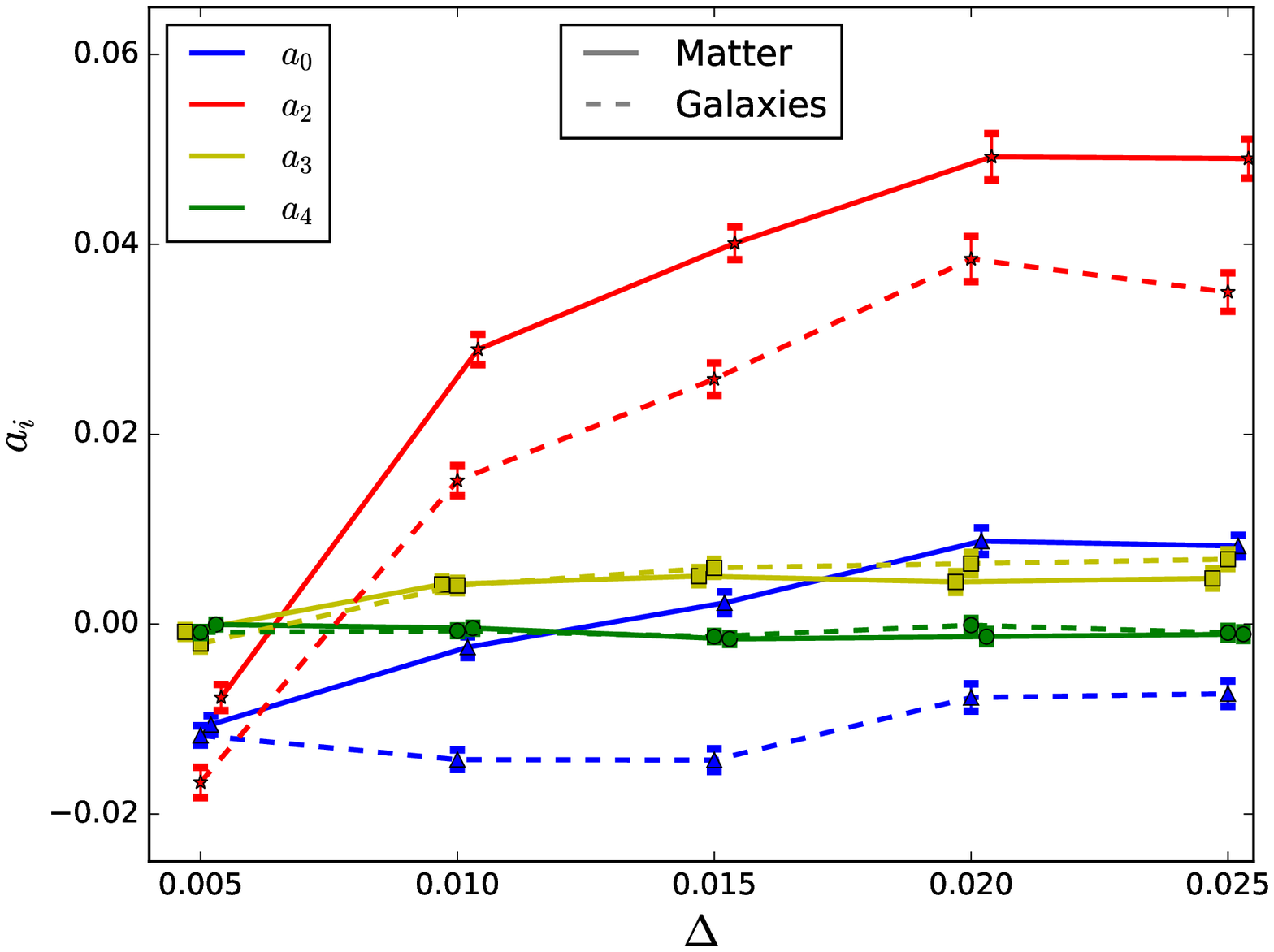}
  \caption{(Top Panel) The shape parameters $a_{0,2,3,4}$ as a function of $\theta_{\rm G}$ in arcminutes, for our fiducial redshift shells $z_{\rm min}=0.25$ and $\Delta=0.02$. We exhibit $a_{0,2,3,4}$ as blue triangles, red stars, yellow squares, and green circles respectively. Solid/dashed lines denote the dark matter/galaxy data. The $a_{0,2}$ terms are the most significant contributors to the distortion from a purely Gaussian genus curve, and $a_{3,4}$ constitute negligible effects. (Bottom Panel) the same quantities $a_{0,2,3,4}$ now as a function of $\Delta$, fixing $\theta_{\rm G} = 27.5 \, {\rm arcmin}$. The same labeling scheme applies. One can see that the shape parameters are relatively insensitive to $\Delta$ for $\Delta > 0.015$, and $a_{0,2}$ are the dominant contributors to departures from the Gaussian form. There is a sign change in $a_{2}$ as we transition to smaller scales.}
  \label{fig:4}
\end{figure}

We plot the shape parameters $a_{0,2,3,4}$ as a function of $\theta_{\rm G}$ (top panel, $\Delta = 0.02$) and $\Delta$ (bottom panel, $\theta_{\rm G} = 27.5 \, {\rm arcmin}$) in Figure\,\ref{fig:4}. Once again, the solid/dashed lines represent dark matter and mock galaxy samples respectively. The $a_{0,2,3,4}$ parameters describe departures of the shape of the genus curve away from its Gaussian form. 

One can see that the genus curve is essentially insensitive to $a_{3,4}$ even on very small angular scales. The only significant shape modifiers are related to the shift parameter $a_{0}$ and the quadratic term $a_{2}$. This is in agreement with semi-analytic results \citep{2000astro.ph..6269M} when one adopts the $\nu_{\rm A}$ parameterization. The $a_{2}$ term changes sign when we take thin slices of the field. For negative $a_{2} < 0$, the genus exhibits a so-called `meatball shift' to the left of the origin $\nu_{\rm A} = 0$. This effect was observed in \citep{1989ApJ...345..618M} and is reproduced here for $\Delta < 0.005$. For thick slices, in which structures are stacked, one finds $a_{2} > 0$ which corresponds to a shift of the genus curve to the right of the origin. This suggests a preponderance of `bubbles' in the density field. The scale at which this transition occurs will depend on the correlation length of the point distribution. For the dark matter field, the transition occurs at smaller $\Delta$. 

For thick slices the parameters $a_{0,2,3,4}$ remain relatively insensitive to $\Delta$. We also observe a slow decrease in $a_{0,2}$ as we increase the angular smoothing scale $\theta_{\rm G}$. This corresponds to an approach to the Gaussian limit. This occurs for both dark matter and simulated galaxies. The dark matter data exhibits stronger non-Gaussian behavior than the galaxies at all scales probed.

\begin{figure}
  \includegraphics[width=0.45\textwidth]{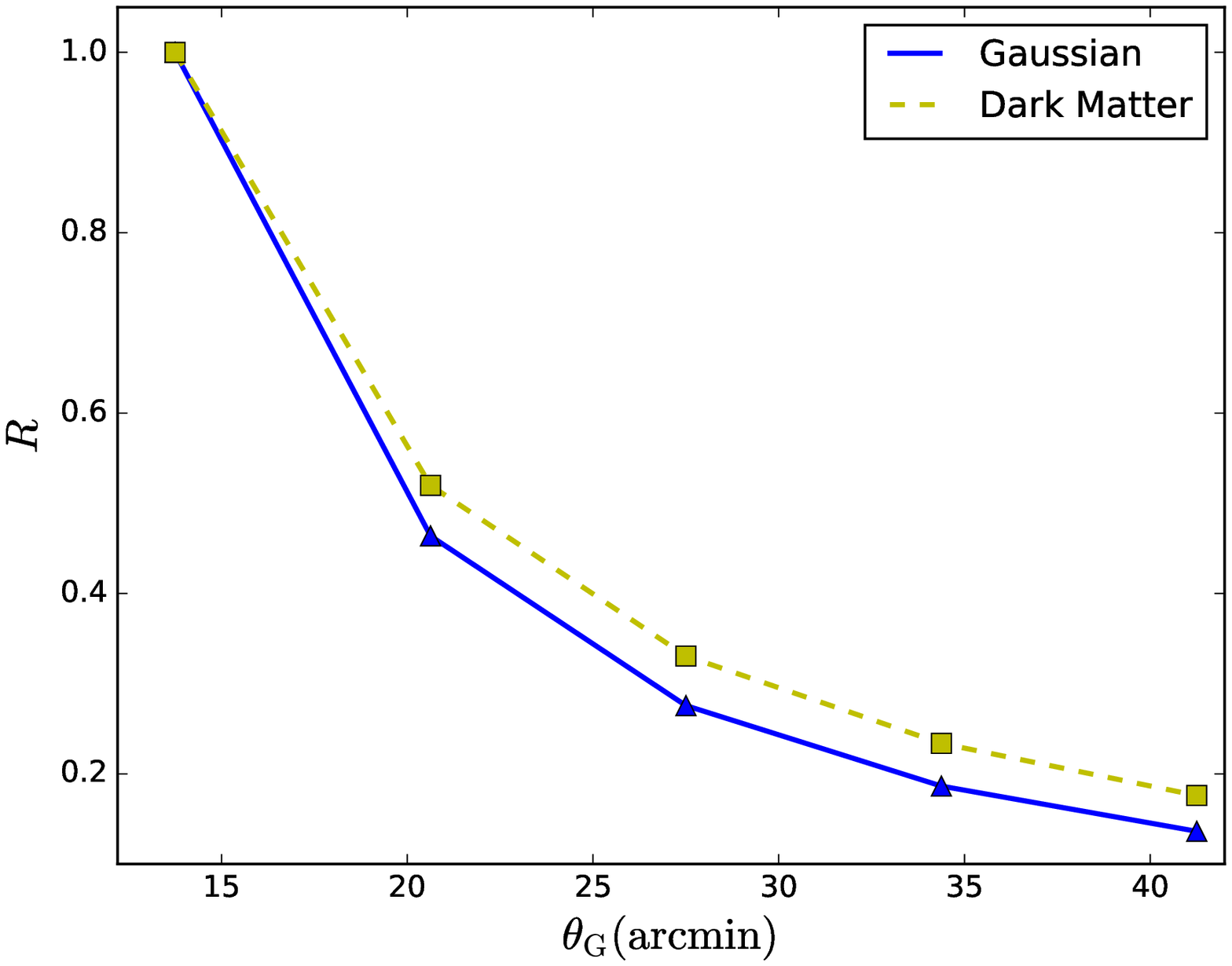}\\
  \includegraphics[width=0.45\textwidth]{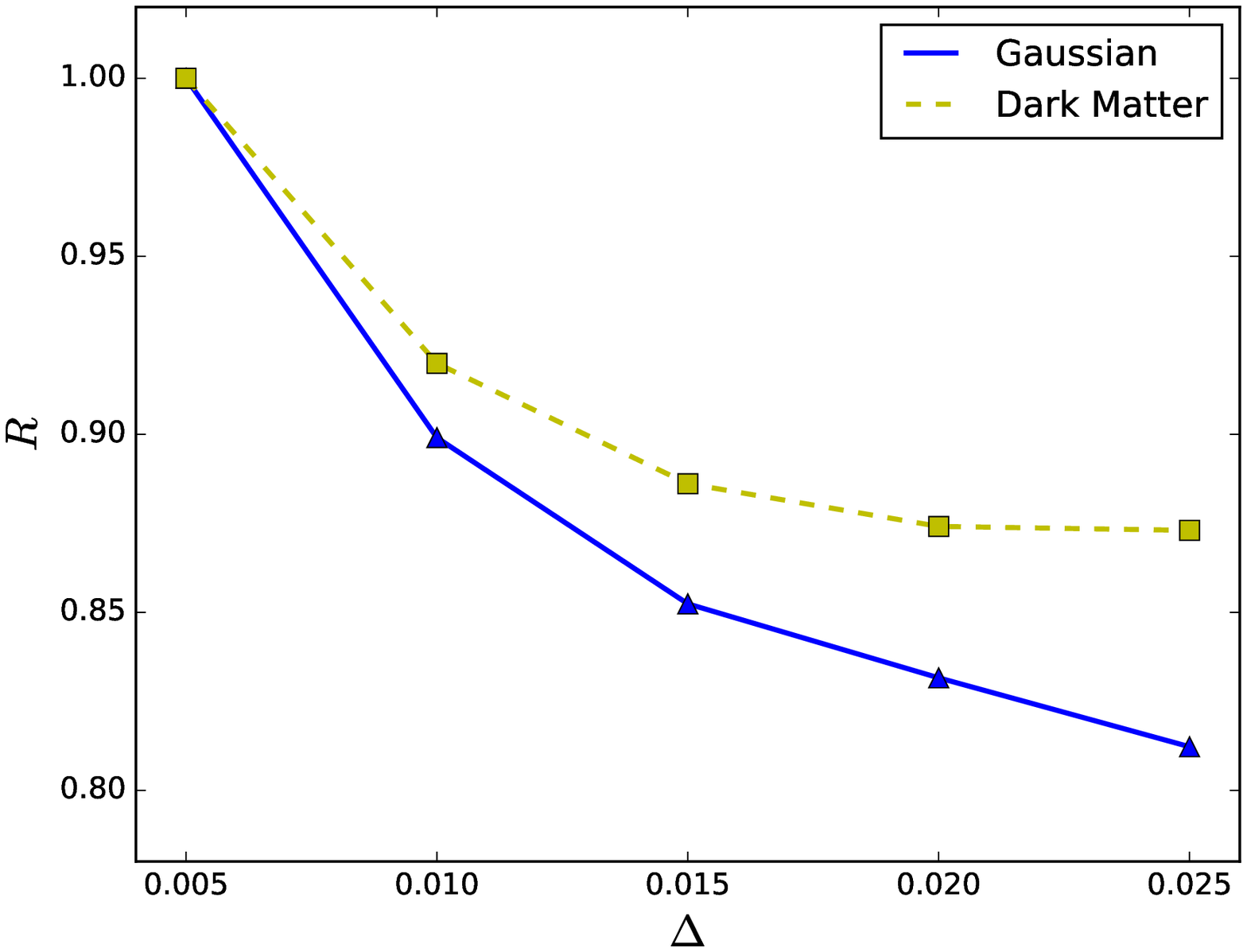}
  \caption{(Top Panel) The statistic $R=a_{1}(\theta_{\rm G},\Delta=0.002)/a_{1}(\theta_{\rm G}=13.75\, {\rm arcmin},\Delta=0.002)$ as a function of $\theta_{\rm G}$ for a Gaussian field (blue solid) and the HR4 $z=0$ dark matter field (yellow dashed). The Gaussian case decreases as $\theta_{\rm G}^{-2}$ as expected; however the nonlinear field exhibits a more shallow dependence on the smoothing scale. (Bottom panel) The $R$ statistic as a function of $\Delta$ for a Gaussian field (blue solid) and the HR4 $z=0$ dark matter field (yellow dashed). Once again, the nonlinear field exhibits less-steep dependence on the smoothing scale. The $\Delta$ dependence becomes increasingly negligible in the thick shell limit $\Delta > 0.015$. }
  \label{fig:R}
\end{figure}

The statistic ($\ref{eq:dg_fid}$) informs us of departures of the measured genus from Gaussianity. To study only the effect of $\theta_{\rm G},\Delta$ on the genus curve, we define the function 

\begin{equation} R = {a_{1}(\theta_{\rm G},\Delta) \over a_{1}(\theta_{0},\Delta_{0})} , \end{equation}

\noindent where $a_{1}$ is the amplitude of the genus and $\theta_{0}, \Delta_{0}$ are arbitrary values of our smoothing parameters that conveniently normalize $R$. We calculate $R$ for both a Gaussian field and the Horizon Run 4 dark matter data by fixing $\Delta_{0} = 0.005$, $\theta_{0}=13.75 \, {\rm arcmin}$ and varying $\theta_{\rm G}$, and subsequently $\Delta$. We exhibit $R$ in Figure \ref{fig:R} as a function of $\theta_{\rm G}$ (top panel) and $\Delta$ (bottom panel). We calculate its value for both a Gaussian field (blue solid line) and the HR4 $z=0$ dark matter field (yellow dashed line). One can observe the expected $R \sim \theta_{\rm G}^{-2}$ behavior for the Gaussian field, and a less-steep dependence for the nonlinear data. The dark matter field is quite insensitive to $\Delta$ in the thick shell limit $\Delta > 0.015$.

%%%%%%%%%%%% FIGURES %%%%%%%%%%%%%%%%%%%%%%%%%%%%%%%%%%%

%%%%%%%%%%%% FIGURES %%%%%%%%%%%%%%%%%%%%%%%%%%%%%%%%%%%

\section{Non-Linear effects}

The genus of a Gaussian density field has a particularly simple form, and all information is encoded in the amplitude of the function. However, the matter density in the low redshift universe is highly nonlinear and non-Gaussian. Numerous complications arise that can modify both the shape and amplitude of the genus curve. In what follows, we discuss four dominant effects: finite size pixels, gravitational evolution, RSD and shot noise/galaxy bias.

\subsection{Pixel Effects}
\label{sec:pix}

The first departure from Gaussianity considered is an entirely numerical artifact. A pixelated density field $\delta_{ijk}$ will perfectly represent its continuous counterpart $\delta(x,y,z)$ only in the limit $p \to 0$, where $p$ is the typical size of a pixel. Pixels of finite size will introduce spurious modifications to the genus of order ${\cal O}(p^{2} \langle k^{2} \rangle)$. Analytic forms exist that correct the genus for finite pixel effects \citep{1989ApJ...345..618M}; however, the form of the correction depends on both the shape of the pixel and the mass binning scheme. Here, we fit the effect using the Hermite expansion ($\ref{eq:fit}$). We take a set of smoothing scales $\theta_{\rm G}$ and for each calculate the genus using three different pixel sizes $N_{\rm s} = (1024, 512, 256)$, where $N_{\rm pixel}=12\times N_{\rm s}^{2}$. We then define 

\begin{equation}\label{eq:dg_pix} \Delta g_{\rm 2D, pix} = g_{\rm 2D}(\theta_{\rm G},N_{\rm s}) - g_{\rm 2D}(\theta_{\rm G},N_{\rm s}=1024) , \end{equation} 

\noindent where $g_{\rm 2D}(\theta_{\rm G},N_{\rm s})$ are the values of the genus extracted from the data for $N_{\rm s}=(512,256)$. We fit the expansion ($\ref{eq:fit}$) to ($\ref{eq:dg_pix}$) - If we assume that the finest pixelation of the sphere, $N_{\rm s} = 1024$, provides an accurate representation of the true density field, then the coefficients $a_{0-4}$ will inform us of the dependence of the genus curve on the size of the pixels.

\begin{figure}
  \includegraphics[width=0.45\textwidth]{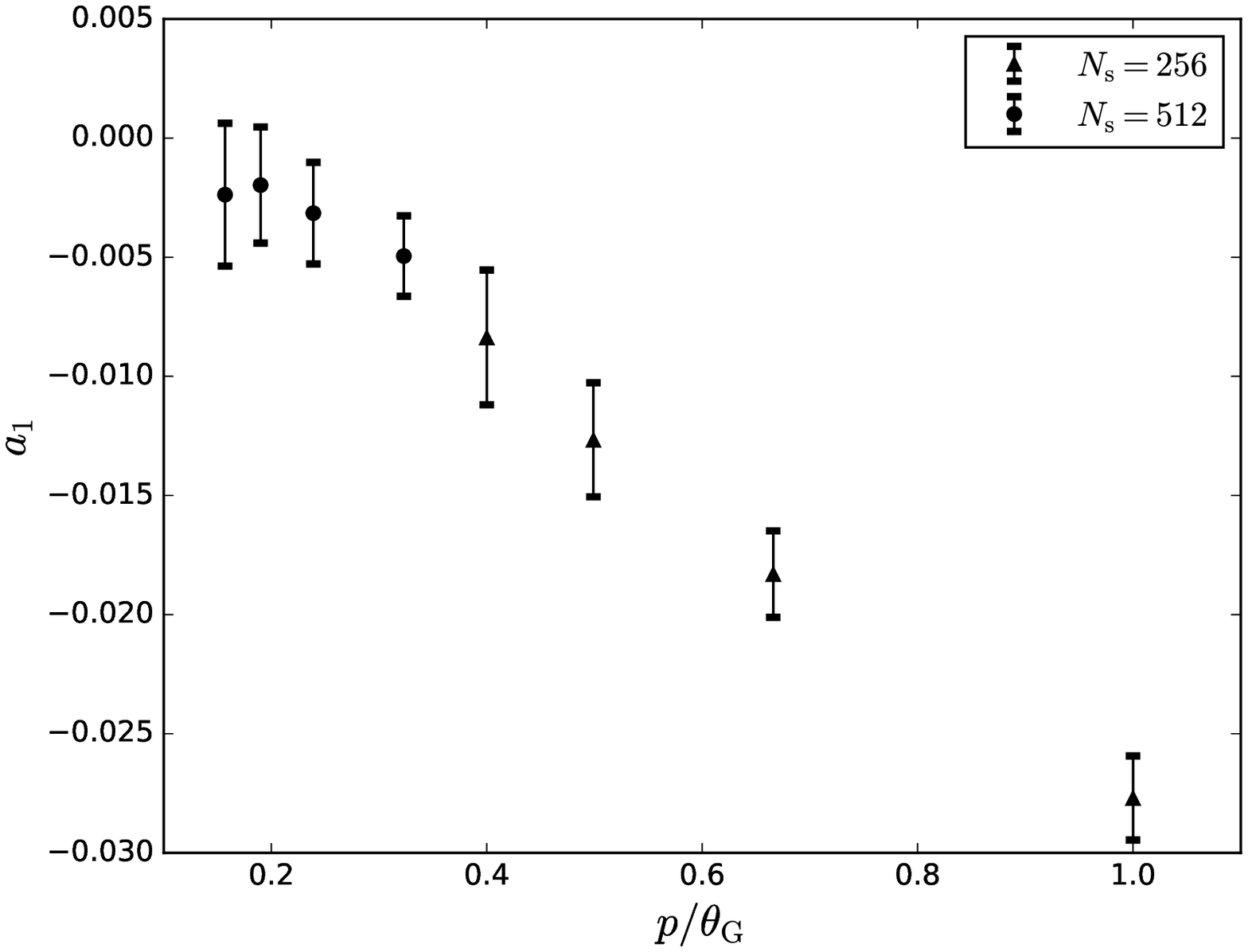}\\
  \includegraphics[width=0.45\textwidth]{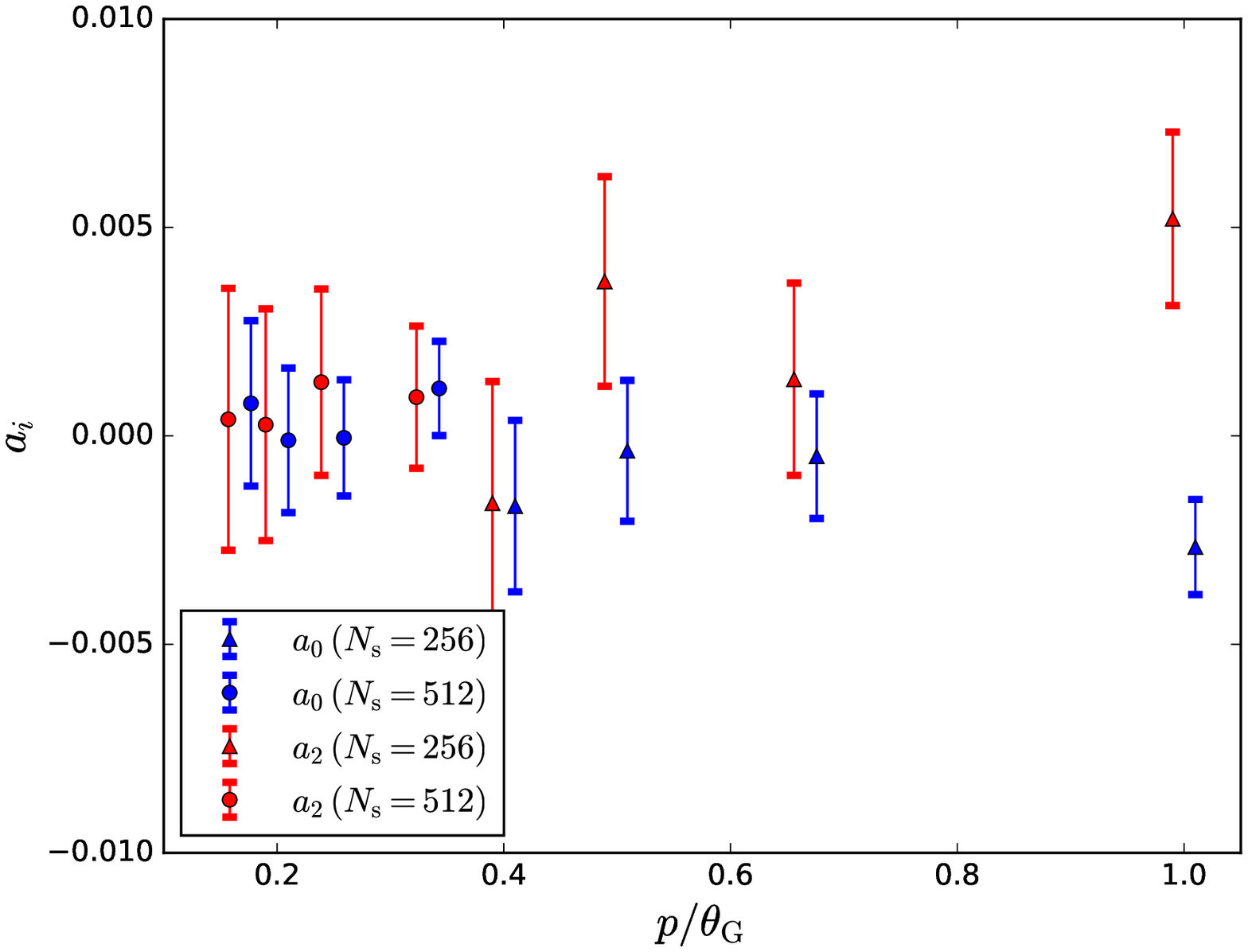}\\
  \caption{(Top panel) The Hermite polynomial coefficient $a_{1}$ as a function of pixel size $p/\theta_{\rm G}$. Here $a_{1}$ describes the finite pixel effect on the amplitude of the genus, as explained in the text. The black circles have been generated using pixel size $N_{\rm s} =512$ and black triangles $N_{\rm s} = 256$. One can observe a systematic decrease in the amplitude as we increase the pixel size. Pixel effects are sub-percent level for $p < \theta_{\rm G}/3$. (Bottom panel) The Hermite coefficient parameters $a_{0},a_{2}$ (blue and red points respectively), as a function of pixel size $p/\theta_{\rm G}$. Triangles and circles represent $N_{\rm s}=256$ and $N_{\rm s}=512$. The $a_{0,2}$ parameters are consistent with zero over the range considered, indicating that pixel size effects do not significantly modify the shape of the genus curve.}
  \label{fig:6}
\end{figure}

In Figure\,\ref{fig:6} (top panel) we exhibit the coefficient $a_{\rm 1}$ as a function of $p/\theta_{\rm G}$. We can consider this curve as representative of the magnitude of the finite pixel effect on the amplitude of the genus as a function of $p/\theta_{\rm G}$. The shape of the pixels in the HEALPix projection are not standardized, so we define $p = (10800/\pi)\sqrt{4\pi/N_{\rm pixel}}$ (in arcminutes). We use the mock galaxy data in this section. 

The term $a_{1}$ decreases with $p/\theta_{\rm G}$. The pixel effect is $\sim 1.2\%$ when $p/\theta_{\rm G} \sim 0.5$ and $\sim 3\%$ at $p/\theta_{\rm G} \sim 1$. One should choose $p < \theta_{\rm G}/3$ to safely ensure that pixelation effects are less than $1\%$. 

The Hermite polynomial coefficients $a_{0,2}$ are exhibited in the bottom panel of Figure \ref{fig:6}. These quantities do not vary appreciably with pixel size for $p/\theta_{\rm G} < 1$, although there is increasing scatter for large pixels. For $p < \theta_{\rm G}/3$, no spurious shape dependence will arise due to finite size pixels.

%%%%%%%%%%%%%%%%%%%% FIGURES %%%%%%%%%%%%%%%%%%%%%%%%

%%%%%%%%%%%%%%%%%%%%%%%%%%%% GRAVITATIONAL CLUSTERING %%%%%%%%%%%%%%%%%%%%%%%%%%

\subsection{Gravitational Clustering}
\label{sec:GC}

As the density field evolves from high redshift, it undergoes collapse into knots and filaments. The process of gravitational evolution will significantly modify the morphology of structures (on small scales and at late times, where the field is strongly nonlinear). The effect of gravitational evolution on the power spectrum of the dark matter density field is to transfer power from large to small scales. As the amplitude of the genus is related to an integral with the power spectrum acting as a window function (see equation ($\ref{eq:gg}$)), one might expect the amplitude to increase from high to low redshift. However, the effect of gravitational smoothing will decrease the amplitude of a gravitationally evolved field relative to a Gaussian field drawn from the same power spectrum. Furthermore, when averaging the density field over suitably large scales, the amplitude should exhibit no appreciable evolution, as the linear power spectrum preserves its shape with time and the number of structures will be conserved. In this section, we use Horizon Run 4 dark matter data to study the evolution of the genus with redshift.

We take the dark matter snapshot data at five redshifts $z=(0, 0.3, 0.7, 1.0, 4.0)$ and calculate the genus. Specifically, we place eight observers in the simulation box at each epoch and calculate the genus curve in shells located at distance $d_{\rm cm} = 738 \, {\rm Mpc/h}$ from the observer. We fix $(\theta_{\rm G}, \Delta)$ to their fiducial values in this section.

Our comparison of snapshot data does not accurately reflect the realities of observational cosmology, where one would use lightcone data and galaxies as opposed to dark matter. However, in this work we wish to disentangle various systematics that act to modify the genus curve. As such, we measure the evolution of the genus using dark matter particles; in doing so we largely avoid shot noise and galaxy bias that are present in galaxy catalogs. We consider these systematics in section \ref{sec:shot_noise}.

\begin{figure}
  \includegraphics[width=0.45\textwidth]{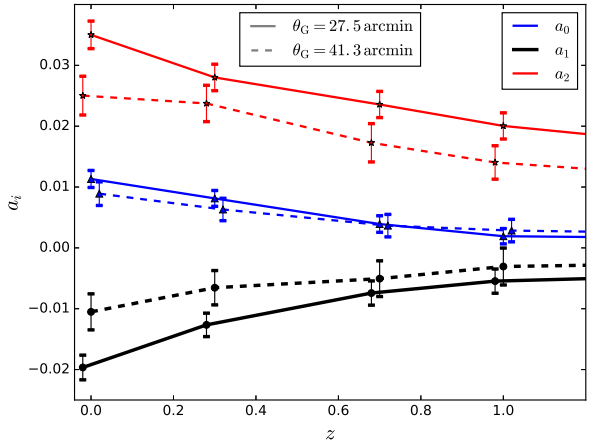}\\
  \caption{The evolution of the Hermite coefficients $a_{0-2}$ with redshift. The solid (dashed) lines represent smoothing scales $\theta_{\rm G} = 27.5, (48.1) \, {\rm arcmin}$ respectively. The genus amplitude, represented by $a_{1}$, decreases as redshift decreases due to gravitational smoothing - a Gaussian field will maximize the genus amplitude of a given power spectrum. This effect is mitigated as we increase the smoothing scale. The shape parameters $a_{0,2}$ also exhibit considerable redshift evolution, and all quantities approach zero at early times, indicating an approach to Gaussianity.}
  \label{fig:7}
\end{figure}

After calculating the genus in eight shells at each redshift, we construct the difference 

\begin{equation} \label{eq:dz} \Delta g_{\rm 2D, grav}(\nu_{\rm A},z) = g_{\rm 2D}(\nu_{\rm A}, z) - g_{\rm 2D}(\nu_{\rm A}, z_{0}) , \end{equation}

\noindent and fit the low-order Hermite polynomial approximation ($\ref{eq:fit}$) to this function. Here, $z_{0}$ is some arbitrary high redshift, that we take as $z_{0}=4$. The coefficients $a_{0-4}$ will inform us of the evolution of the genus amplitude ($a_{1}$) and shape ($a_{0,2,3,4}$) with redshift. We exhibit the coefficients $a_{0-2}(z)$ in Figure\,\ref{fig:7}. As in Figure\,\ref{fig:4}, $a_{0,2}$ are represented as blue triangles and red stars respectively, and now black lines/points correspond to $a_{1}$. We omit $a_{3,4}$, which are consistent with zero at all redshifts. Solid (dashed) lines denote the choice of smoothing parameters $\theta_{\rm G} = 27.5, (41.3) \, {\rm arcmin}$ respectively, with fixed $\Delta =0.02$. 

When we adopt small-scale smoothing $\theta_{\rm G}=27.5 \, {\rm arcmin}$, one can observe a monotonic increase in the parameters $a_{0,2}$ with decreasing redshift, indicating an increasingly non-Gaussian field at late times. The amplitude decreases slightly over the range $0 < z < 1$, this is due to gravitational smoothing. This effect is mitigated when we use a larger smoothing scale $\theta_{\rm G}=41.3 \, {\rm arcmin}$, and the change in the amplitude of the genus over the range $0 < z < 1$ is less than $\sim 1\%$.

In a companion paper, we will use the amplitude of the genus for cosmological parameter estimation, which requires the genus amplitude to act as a standard ruler. This condition is satisfied only when the density field is smoothed over scales sufficiently large that the number of structures is conserved or when the systematic evolution is well-understood. In Figure\,\ref{fig:7} one can observe clear evolution in the Hermite coefficients $a_{0,1,2}$, which is mitigated when we smooth over larger scales (dashed curves). Because we are probing the nonlinear regime, the parameter $a_{1}$ does not completely specify the form of the genus. Both $a_{0}$ and $a_{2}$ are non-zero and evolve over the range $z\sim (0,1)$, suggesting an evolving shape. The shape evolution will be asymmetric around $\nu_{\rm A} = 0$, as the $H_{0,2}(\nu_{\rm A})$ Hermite polynomials do not respect the fact that the Gaussian genus is an odd function. The $a_{0,2}$ contributions are necessarily redshift-dependent, and should approach their Gaussian expectation value $a_{0,2}=0$ at high redshift and/or large smoothing scales $\theta_{\rm G}$. {The shape parameters $a_{0}, a_{2}$ grow with the variance of the field $\sigma_{0}$.}

\begin{figure}
  \includegraphics[width=0.45\textwidth]{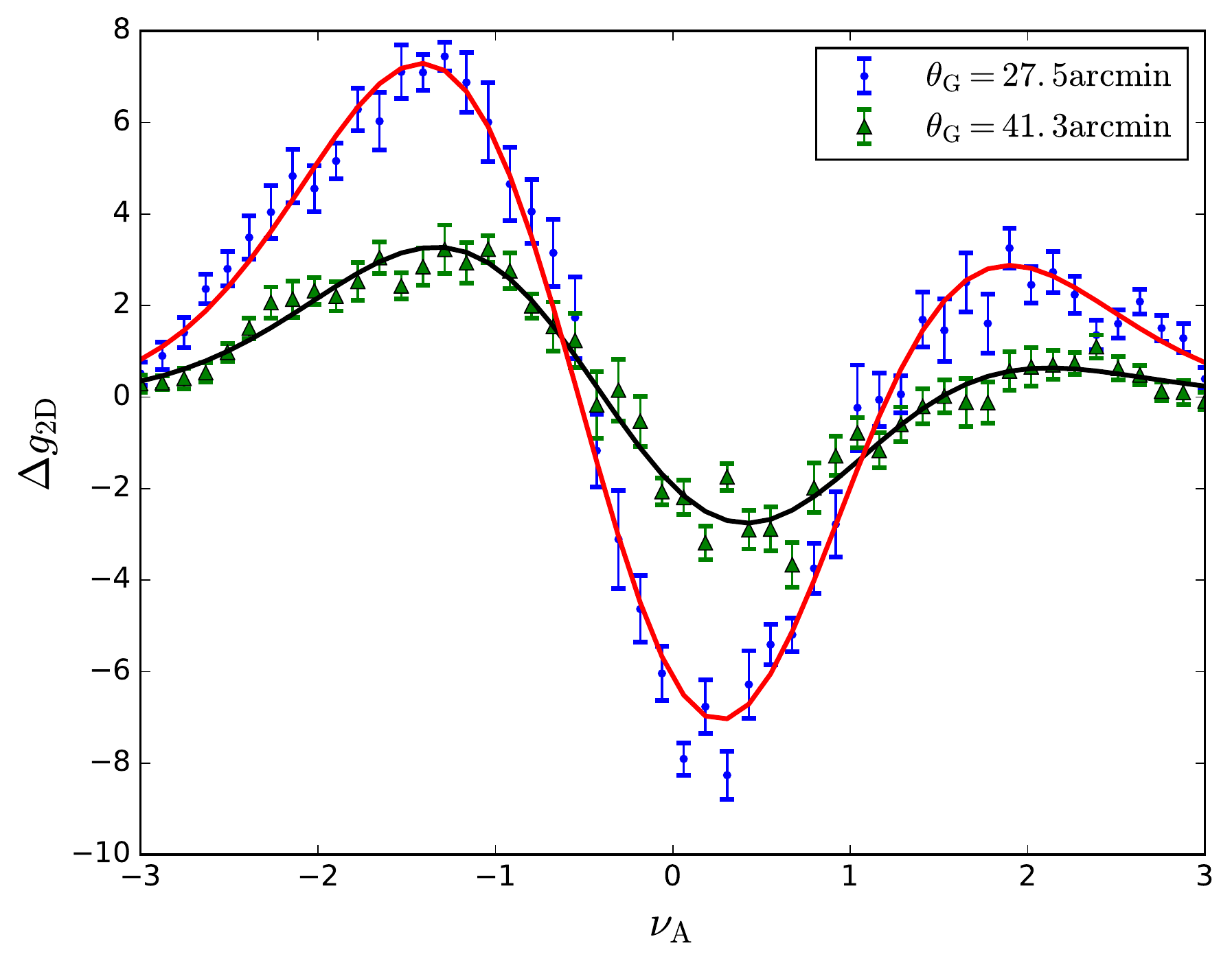}\\
  \caption{The quantity $\Delta g_{\rm 2D, grav}(z,\nu_{\rm A})$ at $z=0$ for two different angular smoothing scales $\theta_{\rm G} = 27.5, 41.3 \, {\rm arcmin}$ (blue and green points). The error bars denote the $\sim 68\%$ uncertainty on the population mean, based on our $n=8$ distinct shells obtained from the simulation box. One observes increasingly nonlinear and asymmetric behavior as we decrease the smoothing scale, with nonlinear effects more apparent at low density. The red and black curves are the best fit to the data sets when we adopt the Hermite expansion ($\ref{eq:fit}$). The curve shape is largely dictated by the asymmetric $a_{0,2}$ parameters.}
  \label{fig:8}
\end{figure}

We exhibit the asymmetry of $\Delta g_{\rm 2D, grav}(z=0)$ for the parameter choices $\theta_{\rm G} = 27.5, 41.3 \, {\rm arcmin}$ in Figure\,\ref{fig:8} (blue and green points), and the best fit theoretical curves are also shown as solid red and black lines, respectively. The error bars represent the $68\%$ uncertainty on the population mean of $\Delta g_{\rm 2D, grav}$ based on the $n=8$ shells within the simulation box. For $\theta_{\rm G} = 41.3 \, {\rm arcmin}$, $\Delta g_{\rm 2D, grav}$ is more symmetric about $\nu_{\rm A}=0$ but as we smooth over successively smaller scales the under-dense region $\nu_{\rm A} < 0$ exhibits an excess in the genus curve relative to $\nu_{\rm A} > 0$, in agreement with three-dimensional results of \cite{Kim:2014axe}. In \cite{2002ApJ...570...44H}, a preponderance of isolated clusters relative to isolated voids has been observed in galaxy data. On the other hand, according to \cite{2005ApJ...633....1P} the shift and asymmetry of the genus curve depend on the smoothing scale and redshift for a given cosmology, as well on the kind of density tracer adopted.

In \cite{2003ApJ...584....1M}, the non-Gaussian signal in the two-dimensional genus was calculated to linear order in $\sigma_{0}$

\begin{eqnarray} \nonumber & &  g_{\rm 2D}(\nu_{\rm A}) = A e^{-\nu^{2}/2} \left[ H_{1}(\nu_{\rm A})+ \left[ {2 \over 3} \left( S^{(1)} - S^{(0)}\right) H_{2}(\nu_{\rm A}) + \right. \right. \\
\label{eq:mat1} & & \qquad \qquad \qquad + \left. \left. {1 \over 3} \left(S^{(2)} - S^{(0)}\right) \right] \sigma_{0} + {\cal O}(\sigma_{0}^{2}) \right] , \end{eqnarray} 

\noindent where the skewness parameters $S^{(0)}, S^{(1)}, S^{(2)}$ are related to the three point statistics of the density field as

\begin{eqnarray} \label{eq:sk1} & & S^{(0)} = {\langle \delta^{3} \rangle \over \sigma_{0}^{4}} , \\ 
\label{eq:sk2} & & S^{(1)} = - {3 \over 4} {\langle \delta^{2} (\nabla^{2} \delta) \rangle \over \sigma_{0}^{2} \sigma_{1}^{2}} , \\ 
\label{eq:sk3} & & S^{(2)} = -3  {\langle (\nabla \delta . \nabla \delta)  (\nabla^{2} \delta) \rangle \over \sigma_{1}^{4}} . \end{eqnarray} 

\noindent We can define $\hat{a}_{0,2}$ coefficients as 

\begin{eqnarray} \label{eq:a0} & & \hat{a}_{0} \equiv {1 \over 3} \left(S^{(2)} - S^{(0)}\right)\sigma_{0}  , \\ 
\label{eq:a2} & & \hat{a}_{2} \equiv {2 \over 3} \left( S^{(1)} - S^{(0)}\right)\sigma_{0} . \end{eqnarray}

The expansion ($\ref{eq:mat1}$) qualitatively agrees with our numerical analysis; throughout this work the $H_{0}(\nu_{\rm A})$ and $H_{2}(\nu_{\rm A})$ Hermite polynomials dominate the non-Gaussian signal. One can test more rigorously the domain in which the expansion ($\ref{eq:mat1}$) matches our numerical results by estimating the three point functions $S^{(0)}$, $S^{(1)}$, $S^{(2)}$ for the density field and directly comparing $\hat{a}_{0,2}$ and $a_{0,2}$. 

We measure the skewness parameters $S^{(0)}, S^{(1)}, S^{(2)}$ from the data. Because we measure both $a_{0,2}$ and $S^{(0)}$, $S^{(1)}$, $S^{(2)}$ from the same simulation snapshot data, we can expect the two approaches to produce statistically consistent $a_{0} \sim \hat{a}_{0}$, $a_{2} \sim \hat{a}_{2}$ coefficients as long as we are within the domain of applicability of the small $\sigma_{0}$ expansion. When the expansion breaks down, we expect to find significant disagreement between the parameters.

\begin{figure}
  \includegraphics[width=0.45\textwidth]{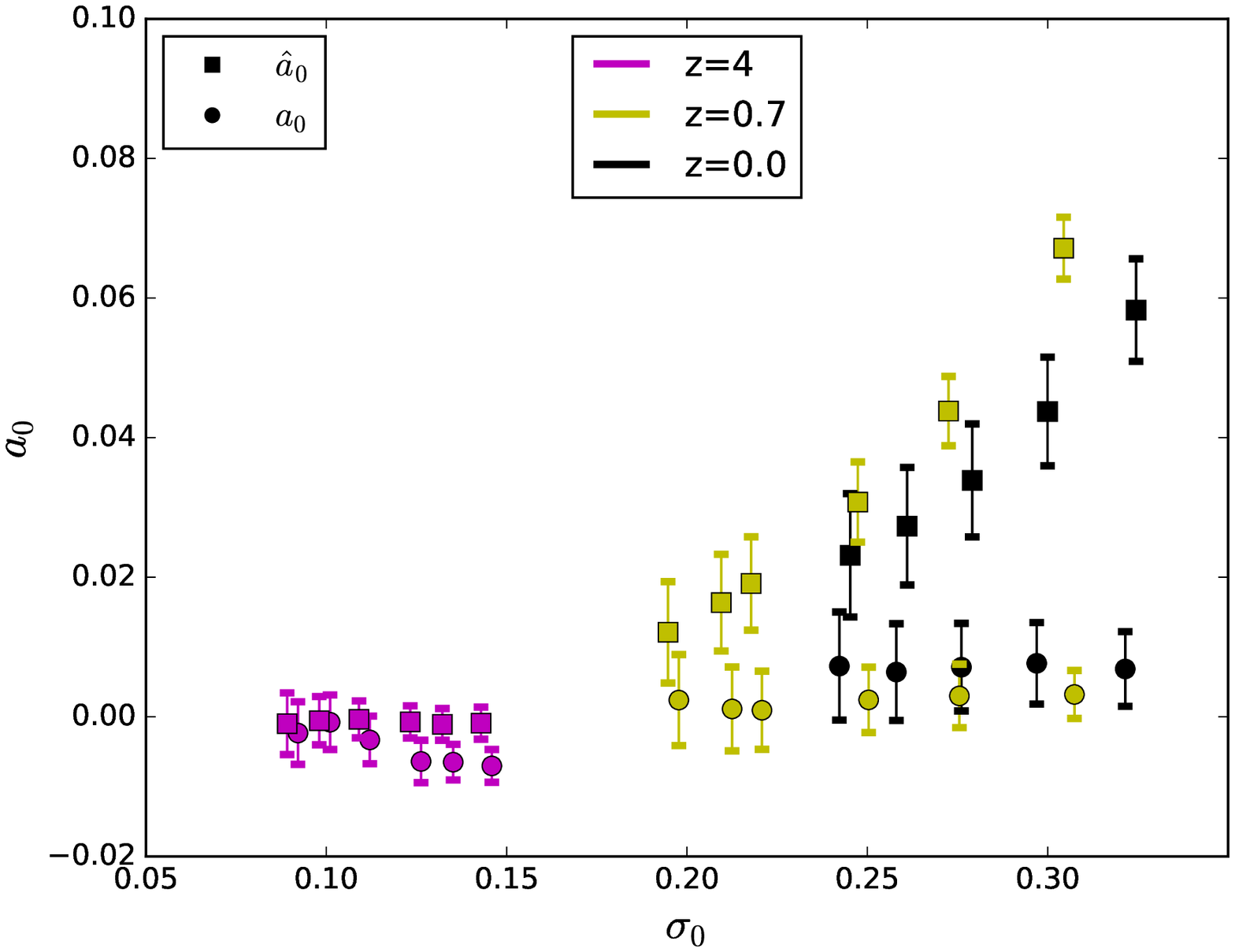}\\
  \includegraphics[width=0.45\textwidth]{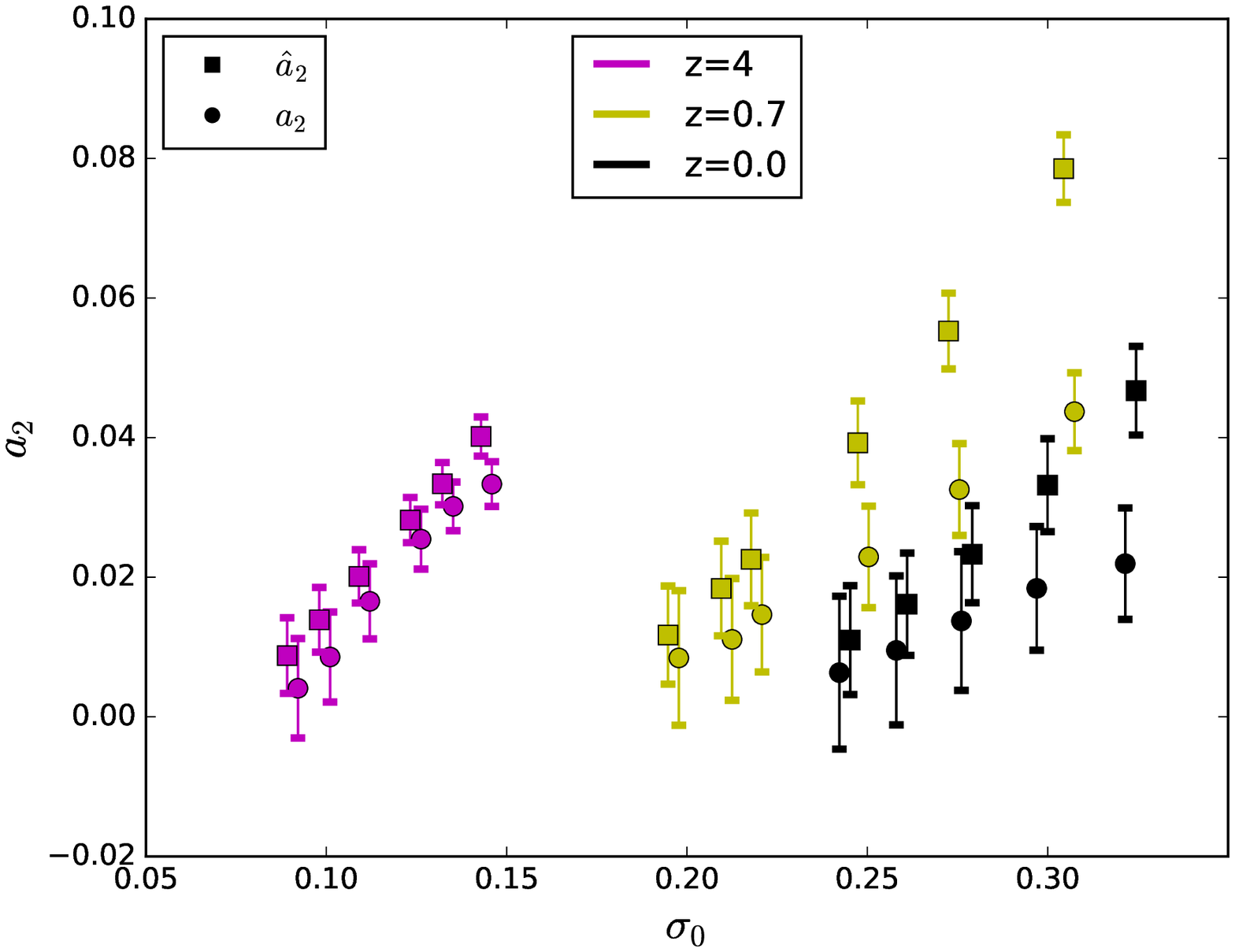}\\
  \caption{(Top panel) The $H_{0}$ Hermite polynomial coefficient obtained by fitting the expansion ($\ref{eq:fit}$) directly to the measured genus curves (circular points) and the theoretical prediction ($\ref{eq:a0}$) (squares), as a function of $\sigma_{0}$. We repeat our analysis for three redshift shells $z=(4,0.7,0)$ (magenta, yellow and black points respectively). The agreement between $a_{0}$ and $\hat{a}_{0}$ is reasonable for small $\sigma_{0}$, with a clear breakdown of the second-order perturbation theory for $\sigma_{0} > 0.2$. (Bottom panel) As in the top panel, but now the parameter $H_{2}$ Hermite polynomial coefficient. One can observe agreement between $a_{2}$ and $\hat{a}_{2}$ to $\sigma_{0} \sim 0.25$, followed by a similar (but less severe) breakdown in the expansion.}
  \label{fig:Mat1}
\end{figure}

In figure \ref{fig:Mat1} we exhibit the $a_{0}, \hat{a}_{0}$ (top panel) and $a_{2}, \hat{a}_{2}$ (bottom panel) parameters as a function of $\sigma_{0}$. We repeat our analysis using three snapshot boxes $z=(0,0.7,4)$ and various smoothing scales. Here, circular points are the $a_{0,2}$ parameters obtained by fitting the expansion ($\ref{eq:fit}$) directly to the genus curves, and squares are calculated using the skewness parameters $S^{(0)}$, $S^{(1)}$, $S^{(2)}$ in the definitions ($\ref{eq:a0},\ref{eq:a2}$). The magenta, yellow, and black points correspond to redshift boxes $z=(4,0.7,0)$ respectively. One can observe agreement between the two methods for small $\sigma_{0}$ in both parameters. The $z=4$ high-redshift field exhibits close agreement between the two methods, with a breakdown at large $\sigma_{0} > 0.2$ for the low-redshift samples. The $H_{0}$ coefficient fails more severely at $\sigma_{0} \sim 0.2$ than $H_{2}$. The $H_{0}$ coefficient is a function of $S^{(2)}$, which is particularly sensitive to small-scale physics, and thus is the most difficult skewness parameter of the three to measure as it involves a higher-order derivatives of the density field, cf. equation ($\ref{eq:sk3}$). The $a_{2}$ parameter is well-measured even at relatively large $\sigma_{0}$ values $\sigma_{0} \sim 0.25$.

We conclude that the genus amplitude does not vary appreciably when the density field is smoothed over sufficiently large scales $R_{\rm G} > 9 {\rm Mpc/h}$, and furthermore that the genus shape parameters provide an unbiased estimate of the third order cumulants of the density field for moderate $\sigma_{0}$ values $\sigma_{0} < 0.2$.

%%%%%%%%%%%%%%%%%%%%%%%%%%%%%%%%% RSD %%%%%%%%%%%%%%%%%%%%%%%%%%%%%%%%%%%%%%%%%%%%%

\subsection{Redshift Space Distortion}

Galaxies possess a peculiar velocity along the line of sight at which we observe them. These motions cause the observed redshift of a galaxy to misrepresent its distance from the observer. The density field in redshift space is therefore distorted relative to its real space equivalent - on large scales, coherent in-fall produces pancake-like distortions, whereas on small scales peculiar velocities of bound objects produce the so-called `finger of God' effect along the line of sight. The peculiar velocities are themselves generated by the gravitational field, leading to a complicated dynamical system. 

Thus far, we have largely focused on dark matter. In this section we choose to study RSD effects on the HR4 simulated galaxy catalog, so that we can apply our results to observational data in future work.

The effect of peculiar velocities on the mock galaxy catalog is straightforward to calculate, as we know both the real and redshift space position of every data point. We therefore calculate the two-dimensional genus in both spaces, and compare them to leading-order theoretical predictions. In real space we simply take a random point in the snapshot box as an observer, and calculate the comoving distance from the observer to each galaxy. We then bin the galaxies into shells and calculate the genus according to section \ref{sec:covgrad}. To obtain the density field in redshift space, we first calculate the redshift corresponding to the comoving distance between galaxy and observer using the correct cosmology. We label this redshift $z_{\rm c}$. We then calculate the peculiar velocity of the galaxy along the line of sight relative to the observer, which we define as $v_{\rm los}$. The actual redshift that would be measured in a galaxy survey is given by 

\begin{equation} 1+z_{\rm obs} = (1+z_{\rm c})\left(1+ {v_{\rm los} \over c} \right) , \end{equation} 

\noindent where $c$ is the speed of light. The redshift space distance to the galaxies is then defined as $d_{\rm rsd}(z_{\rm obs})$. We use $z_{\rm obs}$ to calculate the two-dimensional genus of spherical shells in redshift space.

In \citet{1996ApJ...457...13M}, the effect of RSD on the two-dimensional genus was calculated using linear theory. For a flat two-dimensional slice, if we define the angle between the plane and the line of sight as $\theta_{\rm S}$ then the redshift and real space genus are related as 

\begin{eqnarray} \label{eq:gen_rsd} & & g_{\rm 2D}^{\rm RSD}(\nu,\theta_{\rm S}) = A_{\rm RSD}  g_{\rm 2D}^{\rm real}(\nu) , \end{eqnarray} 

\noindent where 

\begin{equation}\label{eq:amp_rsd} A_{\rm RSD} = {3 \over 2} \sqrt{\left( 1 - {C_{1} \over C_{0}}\right)\left[ 1 - {C_{1} \over C_{0}} + \left({3C_{1} \over C_{0}} -1 \right) \cos^{2}(\theta_{\rm S})\right]}  , \end{equation}

\noindent and 

\begin{equation} {C_{1} \over C_{0}} = {1 \over 3} {1 + (6/5) f b^{-1} + (3/7) (f b^{-1})^{2} \over 1 + (2/3) f b^{-1} +  (1/5) (f b^{-1})^{2}} . \end{equation}

\noindent Here, $b$ is the bias factor, $f = \dot{D}/(HD) \simeq \Omega_{\rm mat}^{6/11}$, and $D$ is the linear growth rate. This result is valid to linear order in the density perturbation $\delta$. One can observe that the effect of RSD on the genus simply corresponds to an amplitude shift, as long as we restrict our analysis to suitably large scales. This calculation is valid for flat slices of the density field; however we expect it to provide a reasonable approximation to our numerical results, as we are choosing thick redshift slices $\Delta =0.02$ and the physical smoothing scale $R_{\rm G}$ on the sphere is much smaller than the observer-shell distance.

\begin{figure}
  \includegraphics[width=0.45\textwidth]{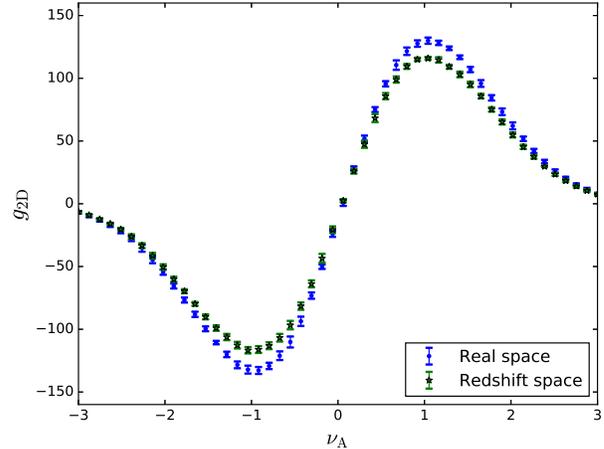}
  \caption{Genus curves in real (blue points) and redshift (green stars) space. One can observe a decrease in the amplitude in redshift space, as predicted from linear theory. Error bars denote the standard deviation of the measurement of eight distinct shells in the snapshot data, and we use mock galaxy data.}
  \label{fig:rsd1}
\end{figure}

We exhibit the genus curves in real (blue error bars) and redshift (green error bars) space in Figure \, \ref{fig:rsd1}. One can observe the amplitude drop in redshift space. If we intend to use the statistic for cosmological parameter estimation, then it is important that we correct for this effect. 

We define a new residual $\Delta g_{\rm 2D, RSD}$ as the difference between the redshift and real space genus curves - 

\begin{equation} \label{eq:rsd_dg} \Delta g_{\rm 2D, RSD}(\nu_{\rm A}) = g_{\rm 2D, RSD}(\nu_{\rm A}) - g_{\rm 2D, real}(\nu_{\rm A}) , \end{equation}

\noindent where $g_{\rm 2D, RSD}(\nu_{\rm A}, \theta_{\rm G}, \Delta), g_{\rm 2D, real}(\nu_{\rm A}, \theta_{\rm G}, \Delta)$ are the genus curves extracted from the $z=0$ snapshot HR4 mock galaxy data in redshift and real space respectively. We fit the Hermite expansion ($\ref{eq:fit}$) to ($\ref{eq:rsd_dg}$). The Hermite coefficient $a_{1}$ will describe the amplitude shift and $a_{0,2,3,4}$ the shape modification as we move between real and redshift space. In the linearized limit, $a_{0,2,3,4}$ should be consistent with zero. Based on our experience in section \ref{sec:RgDel}, we expect both $g_{\rm 2D, RSD}(\nu_{\rm A})$ and $g_{\rm 2D, real}(\nu_{\rm A})$ will depart from the Gaussian limit ($\ref{eq:gg}$). However we make no assumptions regarding the Gaussianity of the field in this section.

Redshift space effects will be sensitive to both $\theta_{\rm G}$ and $\Delta$. For this reason, we repeat our calculation for two different smoothing scales $\theta_{\rm G} = (27.5, 48.1) \, {\rm arcmin}$ and two different shell thickness values $\Delta = (0.005, 0.02)$ (which we call thin and thick shells respectively).

\begin{figure}
  \includegraphics[width=0.45\textwidth]{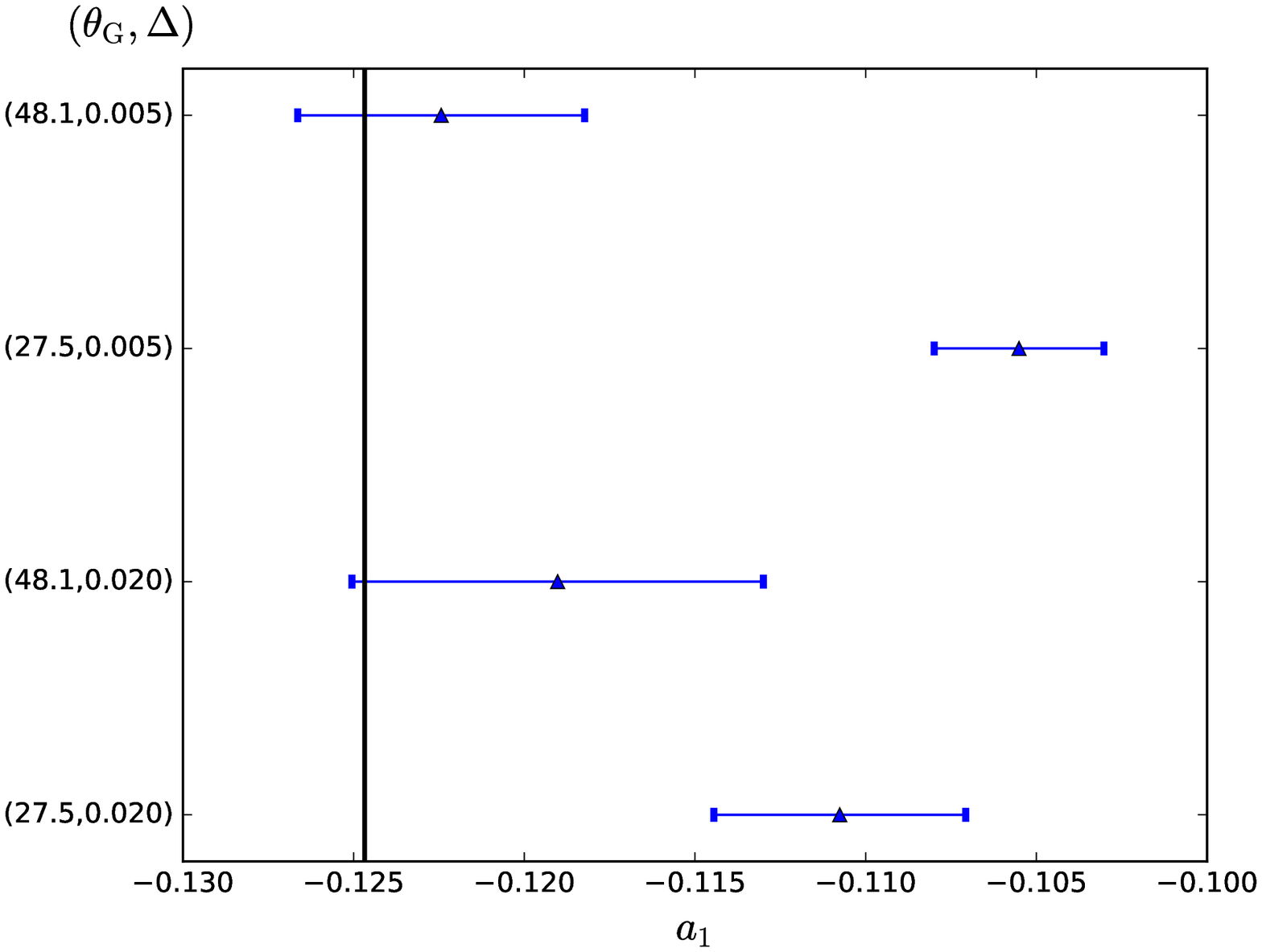}\\
  \includegraphics[width=0.45\textwidth]{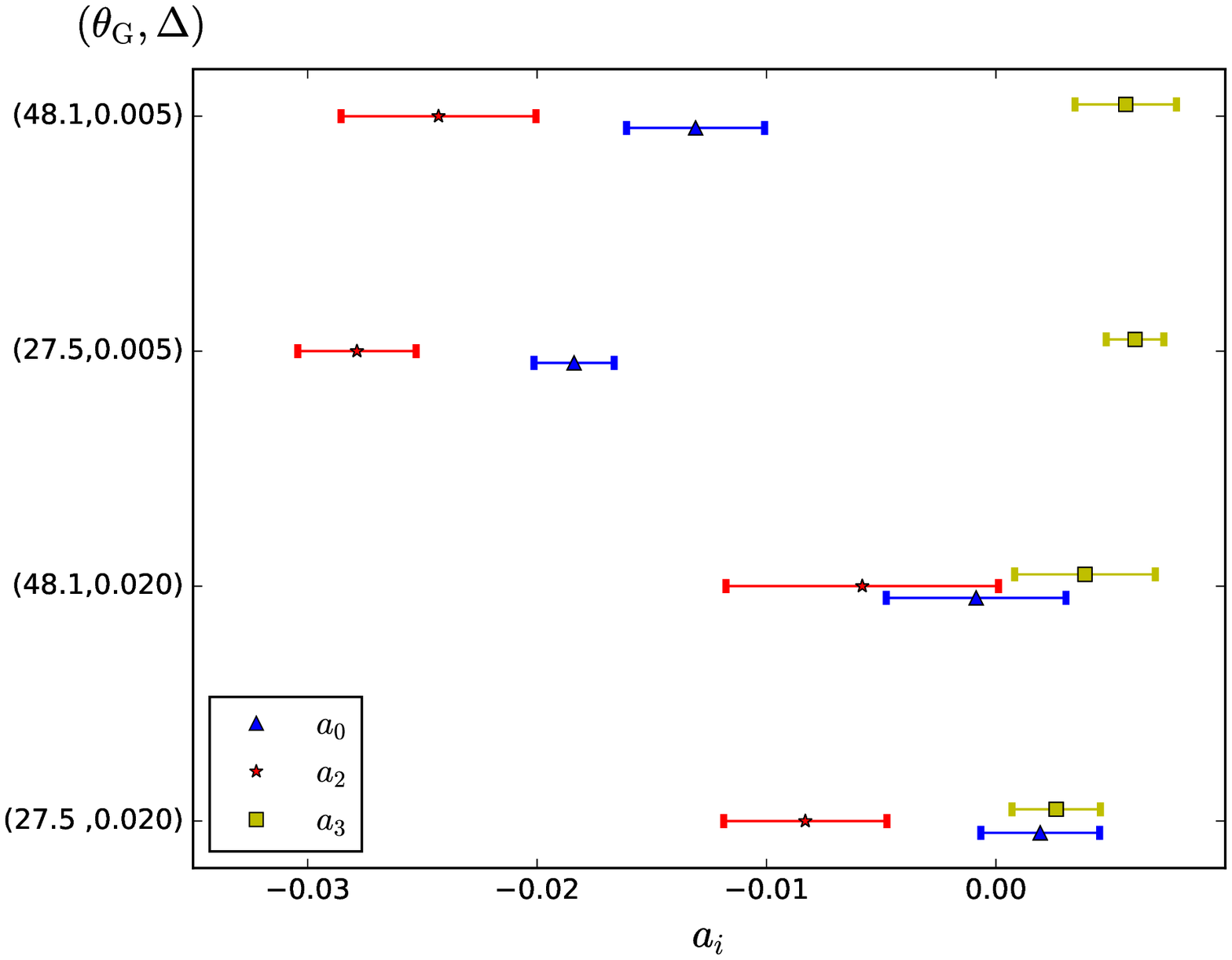}
  \caption{[Top panel] The amplitude coefficient $a_{1}$ for $\Delta g_{\rm 2D, RSD}$ for four sets of smoothing scales $(\theta_{\rm G},\Delta)$. The error bars are the error on the population mean of $g_{\rm 2D,RSD}$, estimated using our $n=8$ shells within the $z=0$ Horizon Run snapshot box. The black vertical line represents the linear approximation ($\ref{eq:amp_rsd}$). The linear approximation correctly approximates the amplitude drop to within $\sim 1.5\%$ for the smoothing scales considered in this work. [Bottom panel] Shape parameters $a_{0,2,3}$ (blue, red, yellow) for the same $(\theta_{\rm G},\Delta)$ smoothing scales. The shape parameters are approximately consistent with zero when using the thick shell $\Delta=0.02$, although the $a_{2}$ parameter exhibits $\sim 1\%$ deviations from $a_{2}=0$.}
  \label{fig:rsd_new}
\end{figure}

In figure\,\ref{fig:rsd_new} (top panel), we exhibit the amplitude coefficient $a_{1}$ for the four possible combinations of smoothing scales $(\theta_{\rm G}, \Delta)$. The linear theory prediction $A_{\rm RSD}-1$ is shown as a solid vertical line, and the error bars are the uncertainty on the mean given our sample of $n=8$ shells. One can observe that the linear prediction is in approximate agreement with our numerical results, to within $\sim 1\%$ when we use $\theta_{\rm G} = 48.1 \, {\rm arcmin}$ smoothing. We have found in previous sections that the genus curve amplitude $a_{1}$ departs significantly from its Gaussian form. Despite this, the amplitude shift between real and redshift space is roughly consistent with the linear prediction ($\ref{eq:amp_rsd}$). However, to use this statistic for cosmological parameter estimation we must improve this agreement to the sub-percent level. 

When using a large smoothing scale $\theta_{\rm G} = 48.1 \, {\rm arcmin}$, the amplitude drop $a_{1}$ is statistically consistent for both thick $\Delta = 0.02$ and thin $\Delta = 0.005$ slices. This is not the case as we decrease the smoothing scale on the sphere to $\theta_{\rm G} = 27.5 \, {\rm arcmin}$, at which point non-linear effects should be accurately modeled. The $\theta_{\rm G} = 27.5 \, {\rm arcmin}$ smoothing cases produce an amplitude shift that is lower than anticipated from linear theory; however, the difference is small. 

In the bottom panel of Figure\,\ref{fig:rsd_new} we exhibit the parameters $a_{0,2,3}$, which inform us how the genus curve shape changes as we pass from real to redshift space. For the thick shell $\Delta = 0.02$, the parameters $a_{0,3}$ are largely consistent with zero; however $a_{2}$ marginally departs from zero for the thick shell $\theta_{\rm G} = 27.5 \, {\rm arcmin}$ case. In the thin shell case, we observe a more severe shape change in all three $a_{0,2,3}$ parameters.

\subsection{Shot Noise and Galaxy Bias}
\label{sec:shot_noise}

\noindent There are two other systematic effects that must be considered when studying the genus: shot noise and galaxy bias. The former is an issue with sampling - as the number density of particles decreases, the point distribution becomes a successively poorer tracer of the underlying density field. The second effect will occur due to the variable relationship between samples of galaxies with different masses and the underlying matter power spectrum. It is not quite trivial to separate these two effects, as the application of mass cuts to the data will also decrease the number density of the catalog.

To study shot noise, we randomly cut dark matter particles from the HR4 sample. After constructing the genus of the cut sample, we compare it to the genus constructed using the full dark matter distribution. To analyze the effect of galaxy bias, we apply mass cuts to the galaxy catalog and compare the resulting genus to the full dark matter equivalent. We take mass cuts such that the random and mass cut catalogs contain the same number of points. The dark matter particle data that we use is an order of magnitude more dense than the galaxies, and so we expect it to accurately represent the underlying dark matter density field. 

Removing galaxies from the sample will generically lead to an artificial increase in the number of structures, which become fragmented by the absence of data. The result will be an increase in the genus amplitude. Furthermore, randomly sampling the galaxies will produce a larger effect than a mass cut, as fragmentation of large overdensities is more likely to occur if we remove high-mass galaxies. 

We define two residual genus functions $\Delta g_{\rm 2D, sn}(\nu_{\rm A},\bar{d})$ and $\Delta g_{\rm 2D, gb}(\nu_{\rm A},\bar{d})$

\begin{eqnarray} & & \Delta g_{\rm 2D, sn}(\nu_{\rm A},\bar{d}) = g_{\rm 2D, mat}(\nu_{\rm A},\bar{d}) - g_{\rm 2D,mat}(\nu_{\rm A},\bar{d}_{0}) , \\
& & \Delta g_{\rm 2D, gb}(\nu_{\rm A},\bar{d}) = g_{\rm 2D, gal}(\nu_{\rm A},\bar{d}) - g_{\rm 2D, mat}(\nu_{\rm A},\bar{d}_{0}) , \end{eqnarray} 

\noindent where $g_{\rm 2D, mat}(\nu_{\rm A}), g_{\rm 2D, gal}(\nu_{\rm A})$ are the genus curves extracted from the dark matter and mock galaxy $z=0$ snapshot data respectively, and $\bar{d}$ is a measure of separation of galaxies on the sphere in arcminutes, defined as $\bar{d} = (10800/\pi)\sqrt{4\pi/N_{\rm p}}$ where $N_{\rm p}$ is the number of dark matter particles or galaxies in the cut catalog. Here, $\bar{d}_{0}$ is the particle separation in the full dark matter sample. When we fit the Hermite expansion ($\ref{eq:fit}$) to these quantities, the resulting Hermite coefficients $a_{0-4}$ will provide a measure of how the genus is affected by sampling effects. 

\begin{figure}
  \includegraphics[width=0.45\textwidth]{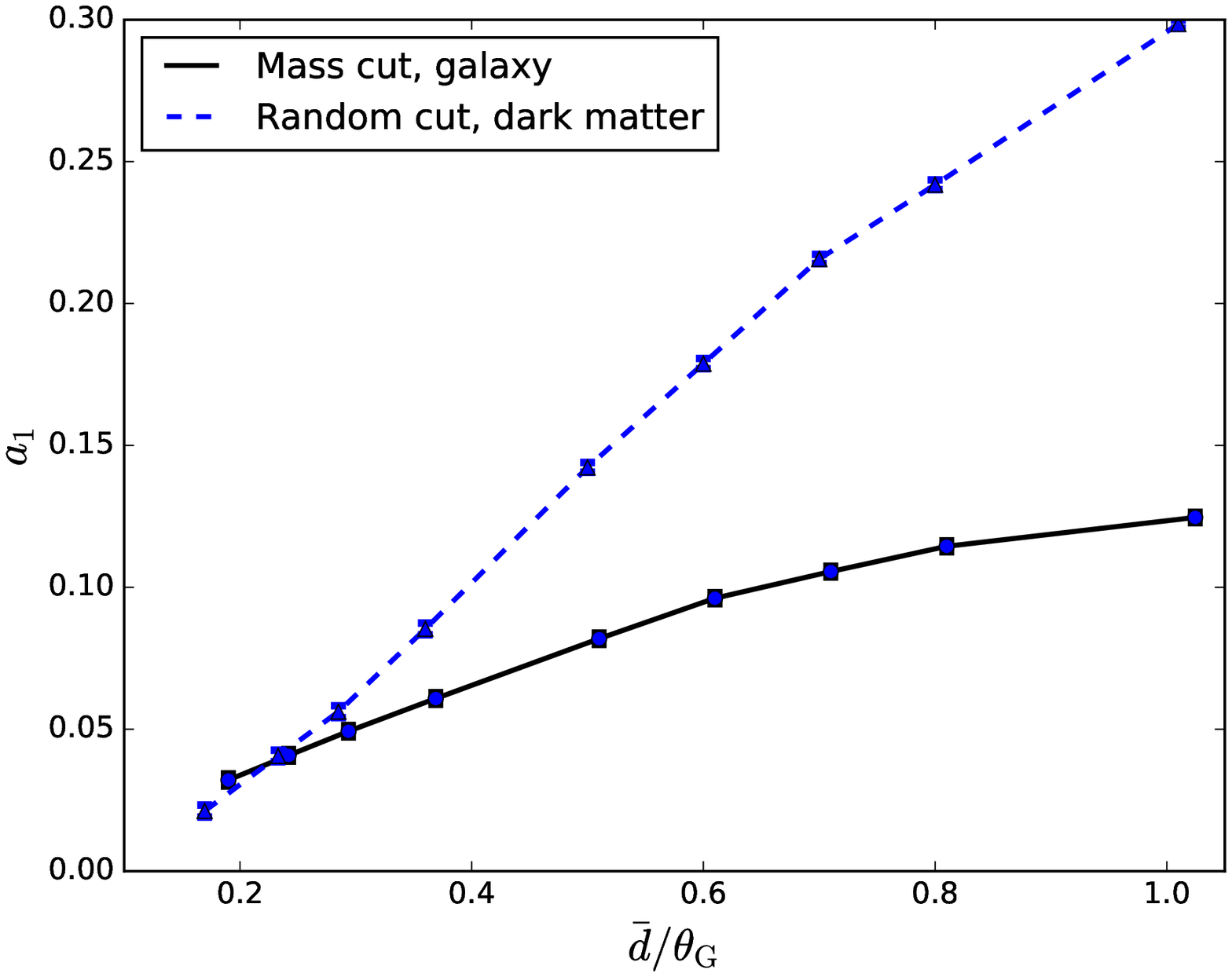}\\
  \includegraphics[width=0.45\textwidth]{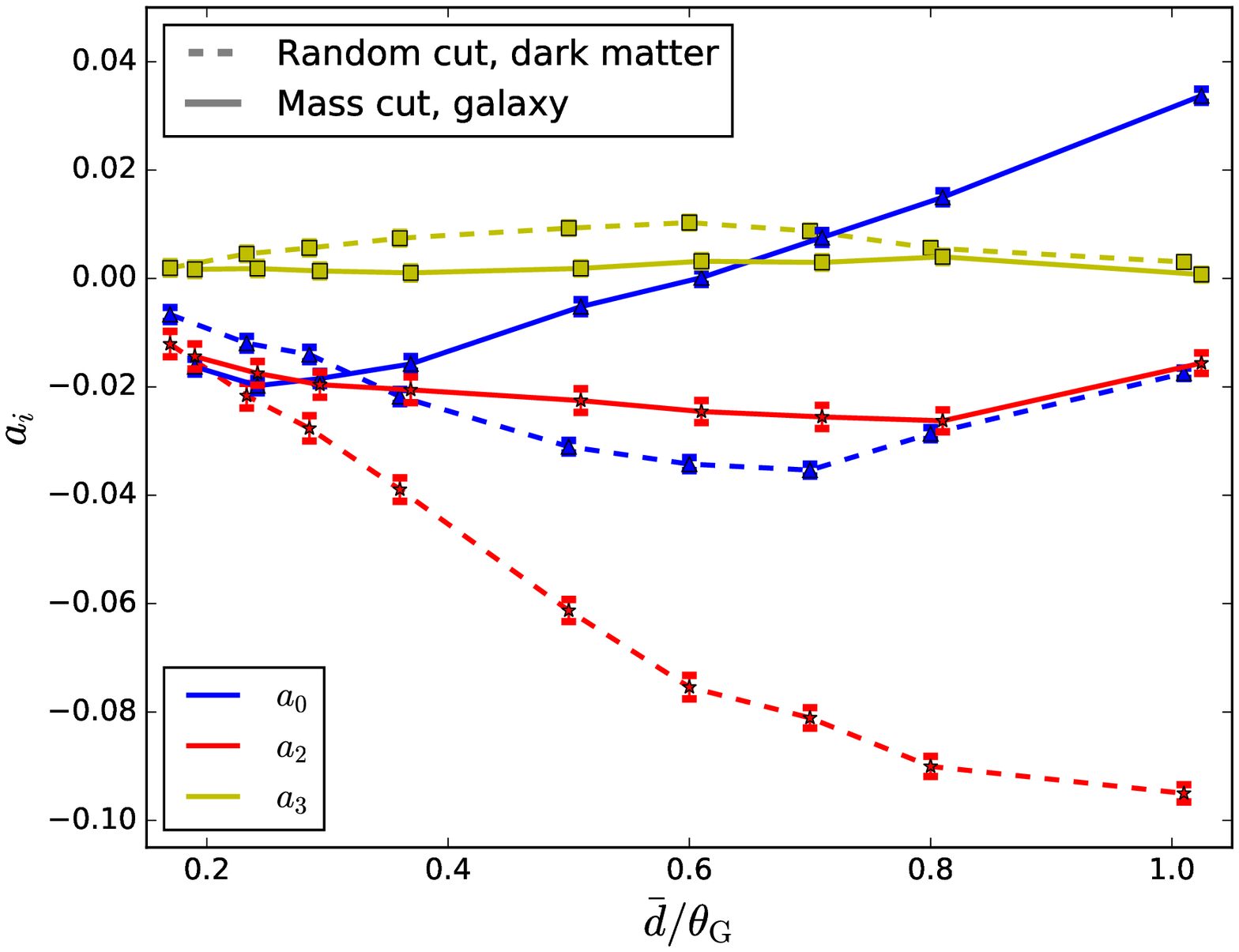}
  \caption{(Top panel) The amplitude coefficient $a_{1}$ as a function of the ratio of the mean separation $\bar{d}$ and smoothing scale $\theta_{\rm G}$. The solid curve represents the galaxy data after successive mass cuts, and the dashed line is a randomly sampled catalog. The randomly sampled catalog exhibits a more extreme departure from the dark matter genus amplitude. (Bottom panel) The shape parameters $a_{0,2,3}$, correspond to blue triangles, red stars and yellow squares. The random and mass cut samples produce statistically equivalent genus curves for $\bar{d}/\theta_{\rm G} < 0.3$. } 
  \label{fig:sn1}
\end{figure}

In Figure\,\ref{fig:sn1} we exhibit the amplitude coefficient $a_{1}$ (top panel) and shape parameters $a_{\rm 0,2,3}$ (bottom panel) as a function of $\bar{d}/\theta_{\rm G}$. As throughout this work, $a_{0,2,3}$ are represented by blue triangles, red stars, and yellow squares. The dashed/solid lines exhibit the effect of making random/mass cuts to the data. We have shifted the mass cut points linearly along the x-axis by a small value for clarity. 

Clearly, the effect of sampling is large - the amplitude $a_{1}$ increases monotonically with $p/\theta_{\rm G}$ when we take a mass cut, and at a greater rate when applying random cuts. This effect is potentially larger than any other considered in this work. As expected, randomly sampling the data produces a more significant effect on the amplitude than applying a mass cut. For a dense sample $\bar{d}/\theta_{\rm G} < 0.2$, the $a_{1}$ amplitude parameters for the random and mass cut catalogs are consistent with one another, within statistical fluctuations. Our choice of sampling is irrelevant, provided our catalog is sufficiently dense.

The shape of the genus curve is also modified as we apply increasingly large cuts to the data, most significantly $a_{2}$. Random sampling again has a more significant effect than applying a mass cut; however, this might be expected given that $a_{0,1,2}$ are correlated. The increasingly negative $a_{2}$ values observed in the bottom panel as we increase $\bar{d}/\theta_{\rm G}$ (red lines) suggest that the field becomes increasingly dominated by `meatball' topology, i.e. a preponderance of isolated clusters. This agrees with our results in section \ref{sec:RgDel} - as the correlation length of the sample increases, the genus has a tendency toward increasingly negative $a_{2}$ values. We observe agreement in the shape parameters between the mass and randomly cut galaxy catalogs for $\bar{d}/\theta_{\rm G} < 0.2$.

%%%%%%%%%%%% PLOTS %%%%%%%%%%%%%%%%%%%%%%%%%%%%

\section{Discussion} 
\label{sec:4}

In this work we have performed an analysis of the properties of the two-dimensional genus of dark matter and mock galaxy catalogs projected onto tomographic redshift shells. We used the latest Horizon Run 4 dark matter simulation and considered effects of gravitational evolution, RSD, shot noise, galaxy bias, and finite size pixels. We draw the following conclusions

\begin{enumerate}
\item{The finite size of pixels modifies the amplitude of the genus curve. However the effect is negligible if we ensure that the pixel size $p$ satisfies $p < \theta_{\rm G}/3$.}
\item{RSD will predominantly modify the genus amplitude. This effect is broadly consistent with linear predictions for smoothing scales $R_{\rm G} \simeq 10 {\rm Mpc/h}$ and thick redshift slices $\Delta \simeq 0.02$, although $\sim 1\%$ discrepancies remain.}
\item{On small scales, the genus amplitude exhibits some evolution with redshift. We find that smoothing on scales $R_{\rm G} \simeq 10 {\rm Mpc/h}$ is sufficient to ensure that the genus amplitude evolves by $\sim 1\%$ over the range $0 < z < 1$}
\item{The effect of galaxy bias is sub-dominant provided the mean separation of galaxies on the sphere satisfies $\bar{d}/\theta_{\rm G} < 0.2$}
\item{The dominant departures in the genus shape from its Gaussian form can be described by the $H_{0,2}$ Hermite polynomials. This agrees with analytic results in the mildly nonlinear regime \citep{2000astro.ph..6269M}.}
\item{The genus amplitude and shape exhibit strong dependence on the number density of the galaxy sample. When comparing the genus at different redshifts, one should make mass cuts to ensure constant number density in each shell.}
\end{enumerate}

Cosmological information is encoded in the amplitude of the genus, so one must be particularly careful to avoid systematic effects when we compare its value in different redshift shells. The most important modifiers of this statistic are RSD and gravitational evolution, as both will explicitly depend upon the cosmological model adopted. If we wish to use the amplitude of the genus for parameter estimation, then these effects are effectively nuisance contributions. Nonlinear finger-of-God effects are mitigated by the stacking procedure that we are forced to adopt when we take $\Delta > 0.02$. We stress that, when dealing with galaxy data, we will not be free to specify the shell thickness - rather it must be larger than the photo-z redshift uncertainty.

We close with a brief discussion on the information content of the genus curve, and the different ways in which one can extract it. Both the shape and amplitude of the genus are related to ratios of cumulants of the density field. For example, the amplitude is a measure of $\sigma_{1}^{2}/\sigma_{0}^{2}$ where $\sigma_{0,1}$ are defined in equations ($\ref{eq:sig0}$) and ($\ref{eq:sig1}$). When the density field is smoothed on physical scales $R_{\rm G} \sim 10 {\rm Mpc/h}$ and the correct cosmology is adopted, the genus amplitude should be conserved as it is measured in successive redshift shells. Both the growth rate and linear bias are canceled when taking the ratio $\sigma_{1}^{2}/\sigma_{0}^{2}$, indicating that the amplitude is insensitive to these quantities. The shape of the angular power spectrum in each redshift shell (when smoothed on large scales) is also relatively insensitive to cosmology. Cosmological parameter dependence enters when we choose an incorrect cosmology, as this will introduce a systematic evolution in our choice of physical smoothing scale $R_{\rm G}$ and also in the surface area of the redshift shell. By choosing an incorrect $R_{\rm G}$ at different redshifts, one is effectively measuring the slope of the power spectrum at different points leading to a systematic increase or decrease of the genus amplitude. Hence, by comparing the genus curve at different redshifts, one can obtain cosmological parameter constraints by minimizing the evolution of the amplitude.

In a series of recent works \cite{Pogosyan:2009rg,Gay:2011wz,Codis:2013exa}, analytic modeling of the shape of the genus curve in the non-Gaussian low redshift universe has been undertaken via an expansion in the variance of the density field. Non-Gaussan shape modifications measure higher-order cumulants of the field. At first order, these are related to the three point functions cf. equations ($\ref{eq:sk1}-\ref{eq:sk3}$). There are two methods by which one can extract information from the non-Gaussian contributions. One is by measuring the higher-order cumulants and comparing them directly with theoretical predictions involving integrals of the Bispectrum, which can be calculated for both $\Lambda$CDM and modified gravity models. The second is by noticing that the non-Gaussian corrections are proportional to powers of $\sigma_{0}$, which will evolve with redshift via the growth factor $D(z)$. Hence by measuring the shape parameters at different redshifts, we can directly reconstruct $D(z)$. 

Clearly, additional information is contained in the higher-order cumulants of the field, which can be extracted using the results of \cite{Codis:2013exa}. One advantage of using the non-Gaussian shape evolution of the genus is that it would allow smoothing on smaller scales, increasing the statistical power of the method. The advantage of using the amplitude is that it provides a purely geometric test of the expansion rate. Because the amplitude and shape information is complementary, a combination of both will optimize constraints on cosmological parameters.

By calculating properties of the two-dimensional density field, we are losing information relative to its full three-dimensional counterpart. However, cosmological information is still contained within $g_{\rm 2D}$ and in a forthcoming publication we will study the constraining power of this statistic when applied to galaxy data. 

\acknowledgements{The authors thank the referee for his helpful comments and also the Korea Institute for Advanced Study for providing computing resources (KIAS Center for Advanced Computation Linux Cluster System) for this work.

This work was supported by the Supercomputing Center/Korea Institute of Science and Technology Information, with supercomputing resources including technical support (KSC-2013-G2-003) and the simulation data were transferred through a high-speed network provided by KREONET/GLORIAD.

Some of the results in this paper have been derived using the HEALPix \citep{Gorski:2004by} package.}

\bibliography{biblio}{}

\begin{thebibliography}{}
\expandafter\ifx\csname natexlab\endcsname\relax\def\natexlab#1{#1}\fi

\bibitem[{Bilicki {et~al.}(2013)Bilicki, Jarrett, Peacock, Cluver, \&
  Steward}]{Bilicki:2013sza}
Bilicki, M., Jarrett, T.~H., Peacock, J.~A., Cluver, M.~E., \& Steward, L.
  2013, arXiv:1311.5246, [ApJS.210,9(2014)]

\bibitem[{Blake {et~al.}(2014)Blake, James, \& Poole}]{Blake:2013noa}
Blake, C., James, J.~B., \& Poole, G.~B. 2014, MNRAS, 437, 2488

\bibitem[{Codis {et~al.}(2013)Codis, Pichon, Pogosyan, Bernardeau, \&
  Matsubara}]{Codis:2013exa}
Codis, S., Pichon, C., Pogosyan, D., Bernardeau, F., \& Matsubara, T. 2013,
  Mon. Not. Roy. Astron. Soc., 435, 531

\bibitem[{{Coles} {et~al.}(1993){Coles}, {Moscardini}, {Plionis}, {Lucchin},
  {Matarrese}, \& {Messina}}]{1993MNRAS.260..572C}
{Coles}, P., {Moscardini}, L., {Plionis}, M., {et~al.} 1993, \mnras, 260, 572

\bibitem[{{Coles} \& {Plionis}(1991)}]{1991MNRAS.250...75C}
{Coles}, P., \& {Plionis}, M. 1991, \mnras, 250, 75

\bibitem[{{Colley}(1997)}]{1997ApJ...489..471C}
{Colley}, W.~N. 1997, \apj, 489, 471

\bibitem[{{Colley} {et~al.}(2000){Colley}, {Gott}, {Weinberg}, {Park}, \&
  {Berlind}}]{2000ApJ...529..795C}
{Colley}, W.~N., {Gott}, J.~R., {Weinberg}, D.~H., {Park}, C., \& {Berlind},
  A.~A. 2000, \apj, 529, 795

\bibitem[{Gay {et~al.}(2012)Gay, Pichon, \& Pogosyan}]{Gay:2011wz}
Gay, C., Pichon, C., \& Pogosyan, D. 2012, Phys. Rev., D85, 023011

\bibitem[{Gorski {et~al.}(2005)Gorski, Hivon, Banday, Wandelt, Hansen,
  Reinecke, \& Bartelman}]{Gorski:2004by}
Gorski, K.~M., Hivon, E., Banday, A.~J., {et~al.} 2005, ApJ., 622, 759

\bibitem[{Gott {et~al.}(2007)Gott, Colley, Park, Park, \&
  Mugnolo}]{Gott:2006za}
Gott, J.~R., Colley, W.~N., Park, C.-G., Park, C., \& Mugnolo, C. 2007, MNRAS,
  377, 1668

\bibitem[{Gott {et~al.}(1986)Gott, Dickinson, \& Melott}]{Gott:1986uz}
Gott, J.~R., Dickinson, M., \& Melott, A.~L. 1986, ApJ., 306, 341

\bibitem[{Gott {et~al.}(1989)Gott, Park, Juszkiewicz, Bies, Bennett, Bouchet,
  \& Stebbins}]{Gott:1989yj}
Gott, J.~R., Park, C., Juszkiewicz, R., {et~al.} 1989

\bibitem[{{Gott} {et~al.}(1987){Gott}, {Weinberg}, \&
  {Melott}}]{1987ApJ...319....1G}
{Gott}, J.~R., {Weinberg}, D.~H., \& {Melott}, A.~L. 1987, \apj, 319, 1

\bibitem[{Gott {et~al.}(1988)}]{Gott:1988rj}
Gott, J.~R., {et~al.} 1988

\bibitem[{Hadwiger(1957)}]{Hadwiger}
Hadwiger, H. 1957, Vorlesungen über Inhalt, Oberfläche und Isoperimetriee
  (Grundlehren der mathematischen Wissenschaften: Springer)

\bibitem[{Hikage {et~al.}(2006)Hikage, Komatsu, \& Matsubara}]{Hikage:2006fe}
Hikage, C., Komatsu, E., \& Matsubara, T. 2006, ApJ., 653, 11

\bibitem[{Hong {et~al.}(2016)Hong, Park, \& Kim}]{Hong:2016hsd}
Hong, S.~E., Park, C., \& Kim, J. 2016, Astrophys. J., 823, 103

\bibitem[{{Hoyle} {et~al.}(2002){Hoyle}, {Vogeley}, \&
  {Gott}}]{2002ApJ...570...44H}
{Hoyle}, F., {Vogeley}, M.~S., \& {Gott}, J.~R. 2002, \apj, 570, 44

\bibitem[{James(2012)}]{James:2011wm}
James, J.~B. 2012, ApJ., 751, 40

\bibitem[{Jiang {et~al.}(2008)Jiang, Jing, Faltenbacher, Lin, \&
  Li}]{Jiang:2007xd}
Jiang, C.~Y., Jing, Y.~P., Faltenbacher, A., Lin, W.~P., \& Li, C. 2008,
  Astrophys. J., 675, 1095

\bibitem[{Kim {et~al.}(2015)Kim, Park, L'Huillier, \& Hong}]{Kim:2015yma}
Kim, J., Park, C., L'Huillier, B., \& Hong, S.~E. 2015, JKAS, 48, 213

\bibitem[{Kim {et~al.}(2014)Kim, Choi, Kim, Kim, Lee, Shin, \&
  Kim}]{Kim:2014axe}
Kim, Y.-R., Choi, Y.-Y., Kim, S.~S., {et~al.} 2014, ApJS., 212, 22

\bibitem[{Laigle {et~al.}(2016)}]{Laigle:2016jxn}
Laigle, C., {et~al.} 2016, arXiv:1604.02350

\bibitem[{Lewis(2013)}]{Lewis:2013hha}
Lewis, A. 2013, Phys. Rev., D87, 103529

\bibitem[{Lewis \& Bridle(2002)}]{Lewis:2002ah}
Lewis, A., \& Bridle, S. 2002, Phys. Rev., D66, 103511

\bibitem[{L'Huillier {et~al.}(2014)L'Huillier, Park, \&
  Kim}]{L'Huillier:2014dpa}
L'Huillier, B., Park, C., \& Kim, J. 2014, New Astron., 30, 79

\bibitem[{Matsubara(1994)}]{Matsubara:1994we}
Matsubara, T. 1994, arXiv:astro-ph/9501076

\bibitem[{{Matsubara}(1996)}]{1996ApJ...457...13M}
{Matsubara}, T. 1996, \apj, 457, 13

\bibitem[{{Matsubara}(2000)}]{2000astro.ph..6269M}
---. 2000, ArXiv Astrophysics e-prints, astro-ph/0006269

\bibitem[{{Matsubara}(2003)}]{2003ApJ...584....1M}
---. 2003, \apj, 584, 1

\bibitem[{Matsubara \& Jain(2001)}]{Matsubara:2000dg}
Matsubara, T., \& Jain, B. 2001, ApJ., 552, L89

\bibitem[{Matsubara \& Suto(1996)}]{Matsubara:1995dv}
Matsubara, T., \& Suto, Y. 1996, ApJ., 460, 51

\bibitem[{{Melott} {et~al.}(1989){Melott}, {Cohen}, {Hamilton}, {Gott}, \&
  {Weinberg}}]{1989ApJ...345..618M}
{Melott}, A.~L., {Cohen}, A.~P., {Hamilton}, A.~J.~S., {Gott}, J.~R., \&
  {Weinberg}, D.~H. 1989, \apj, 345, 618

\bibitem[{{Melott} {et~al.}(1988){Melott}, {Weinberg}, \&
  {Gott}}]{1988ApJ...328...50M}
{Melott}, A.~L., {Weinberg}, D.~H., \& {Gott}, J.~R. 1988, \apj, 328, 50

\bibitem[{{Park} \& {Gott}(1991)}]{1991ApJ...378..457P}
{Park}, C., \& {Gott}, J.~R. 1991, \apj, 378, 457

\bibitem[{{Park} {et~al.}(2001){Park}, {Gott}, \& {Choi}}]{2001ApJ...553...33P}
{Park}, C., {Gott}, J.~R., \& {Choi}, Y.~J. 2001, \apj, 553, 33

\bibitem[{{Park} {et~al.}(1992){Park}, {Gott}, {Melott}, \&
  {Karachentsev}}]{1992ApJ...387....1P}
{Park}, C., {Gott}, J.~R., {Melott}, A.~L., \& {Karachentsev}, I.~D. 1992,
  \apj, 387, 1

\bibitem[{{Park} {et~al.}(2005){Park}, {Kim}, \& {Gott}}]{2005ApJ...633....1P}
{Park}, C., {Kim}, J., \& {Gott}, J.~R. 2005, \apj, 633, 1

\bibitem[{Park \& Kim(2010)}]{Park:2009ja}
Park, C., \& Kim, Y.-R. 2010, ApJ., 715, L185

\bibitem[{Pogosyan {et~al.}(2009)Pogosyan, Gay, \& Pichon}]{Pogosyan:2009rg}
Pogosyan, D., Gay, C., \& Pichon, C. 2009, Phys. Rev., D80, 081301, [Erratum:
  Phys. Rev.D81,129901(2010)]

\bibitem[{Ryden {et~al.}(1989)Ryden, Melott, Craig, Gott, Weinberg, Scherrer,
  Bhavsar, \& Miller}]{Ryden:1988rk}
Ryden, B.~S., Melott, A.~L., Craig, D.~A., {et~al.} 1989, ApJ., 340, 647

\bibitem[{Schmalzing \& Gorski(1998)}]{Schmalzing:1997uc}
Schmalzing, J., \& Gorski, K.~M. 1998, MNRAS, 297, 355

\bibitem[{Schmalzing {et~al.}(1996)Schmalzing, Kerscher, \&
  Buchert}]{Schmalzing:1995qn}
Schmalzing, J., Kerscher, M., \& Buchert, T. 1996, Proc. Int. Sch. Phys. Fermi,
  132, 281

\bibitem[{Speare {et~al.}(2015)Speare, Gott, Kim, \& Park}]{Speare:2013qma}
Speare, R., Gott, J.~R., Kim, J., \& Park, C. 2015, ApJ., 799, 176

\bibitem[{Wang {et~al.}(2012)Wang, Chen, \& Park}]{Wang:2010ug}
Wang, X., Chen, X., \& Park, C. 2012, ApJ., 747, 48

\bibitem[{{Wang} {et~al.}(2015){Wang}, {Park}, {Xu}, {Chen}, \&
  {Kim}}]{2015ApJ...814....6W}
{Wang}, Y., {Park}, C., {Xu}, Y., {Chen}, X., \& {Kim}, J. 2015, \apj, 814, 6

\bibitem[{{Weinberg} {et~al.}(1987){Weinberg}, {Gott}, \&
  {Melott}}]{1987ApJ...321....2W}
{Weinberg}, D.~H., {Gott}, J.~R., \& {Melott}, A.~L. 1987, \apj, 321, 2

\bibitem[{Zunckel {et~al.}(2011)Zunckel, Gott, \& Lunnan}]{Zunckel:2010eh}
Zunckel, C., Gott, J.~R., \& Lunnan, R. 2011, MNRAS, 412, 1402

\end{thebibliography}

\end{document}